\documentclass[article,prb,notitlepage]{revtex4-1}

\usepackage [latin1]{inputenc}
\usepackage{graphicx}
\usepackage{graphics}
\usepackage[usenames]{color}
\usepackage{subfigure}
\usepackage{hyperref}
\usepackage{breakurl}

\newcommand{\la}[1]{\label{#1}}
\newcommand{\Sec}[1]{Sec.~\ref{#1}}

\definecolor{BrickRed}{cmyk}{0,0.89,0.94,0.28}					%%%PANTONE 1805
\definecolor{MidnightBlue}{cmyk}{0.98,0.13,0,0.43}				%%%PANTONE 302
\definecolor{DarkGreen}{rgb}{0.100806,0.495968,0.209979}
\definecolor{orange}{rgb}{0.587167,0.354498,0.146197}


\newcommand{\be}{\begin{equation}}
\newcommand{\ee}{\end{equation}}
\newcommand{\bea}{\begin{eqnarray}}
\newcommand{\eea}{\end{eqnarray}}
\newcommand{\ben}{\begin{enumerate}}
\newcommand{\een}{\end{enumerate}}
\newcommand{\bit}{\begin{itemize}}
\newcommand{\eit}{\end{itemize}}

\newcommand{\Fig}[1]{Fig.~\ref{#1}}

\begin{document}

\title{Portrait of the mathematician as a young man: \\ Revisiting Trinity's Tayler picture}

\author{Alejandro Jenkins}
\email{alejandro.jenkins@ucr.ac.cr}
\affiliation{Laboratorio de F\'isica Te\'orica y Computacional, Escuela de F\'isica, Universidad de Costa Rica, 11501-2060, San Jos\'e, Costa Rica}
\affiliation{International Centre for Theory of Quantum Technologies (ICTQT), University of Gda\'nsk, 80-308, Gda\'nsk, Poland}

\date{February 28, 2026} 

%%%%%%%%%%
%%% ABSTRACT
%%%%%%%%%%
\begin{abstract}
A 17th-century oil painting by an unknown artist, once owned by the Tayler family and now in the collection of Trinity College, Cambridge, is currently identified as a portrait of a young Isaac Barrow.  The sitter is shown pointing to a proposition in Barrow's 1655 edition of Euclid's {\it Elements}, but the portrait bears little resemblance to other depictions of Barrow.  Moreover, Barrow is unlikely to have posed with that book, which appeared in print eight months after he had left England on a four-year tour of the Continent.  Plausible alternatives are that the portrait is of Francis Willughby or Isaac Newton, both of whom resembled the man pictured and may be characterized as disciples of Barrow.  If the Tayler were Newton's portrait, it could shed light on the patronage that allowed him to rise from undergraduate servant ({\it subsizar}) to Lucasian Professor of Mathematics in only five and half years.
\end{abstract}

\maketitle

%\tableofcontents
%
%\newpage

%%%%%%%%%%
%%% INTRODUCTION
%%%%%%%%%%
\section{Introduction: Some puzzles in 17th-century English portraiture}
\la{sec:intro}

Milo Keynes, in his painstaking investigation into the iconography of Isaac Newton, found that the savant sat for fourteen portraits during his lifetime, a record broken in the 18th century only by actor David Garrick.\cite{Keynes-Garrick}  However, in the judgment of art historian Francis Haskell, ``few if any these rise much above the mediocre''.\cite{Haskell}  All but one one of the portraits listed by Keynes were made after Newton was already 59 years old, retired from the university, and settled in London as a senior civil servant.  The exception is the 1689 portrait by Godfrey Kneller, reproduced in \Fig{fig:Newton-Kneller}, which shows a 46-year-old Newton as a Cambridge professor and Member of Parliament, at the height of his intellectual career and in the wake of the publication of his {\it Philosophi\ae~Naturalis Principia Mathematica} (``Mathematical Principles of Natural Philosophy'', 1687).  According to art historian Oliver Millar, it shows Newton ``visibly thinking''.  It is certainly the most interesting of the known portraits of Newton, but it is not a fully polished work.  The background is blank and Newton's gown is only sketched in broad strokes.  According to Newton's biographer Frank Manuel, ``Kneller, who was immensely fashionable and was deluged with commissions, often made quick sketches or painted a head, leaving the rest of the work to one of his numerous assistants.''\cite{Manuel-Kneller}

\begin{figure*} [t]
\begin{center}
	\includegraphics[height=0.62 \textwidth]{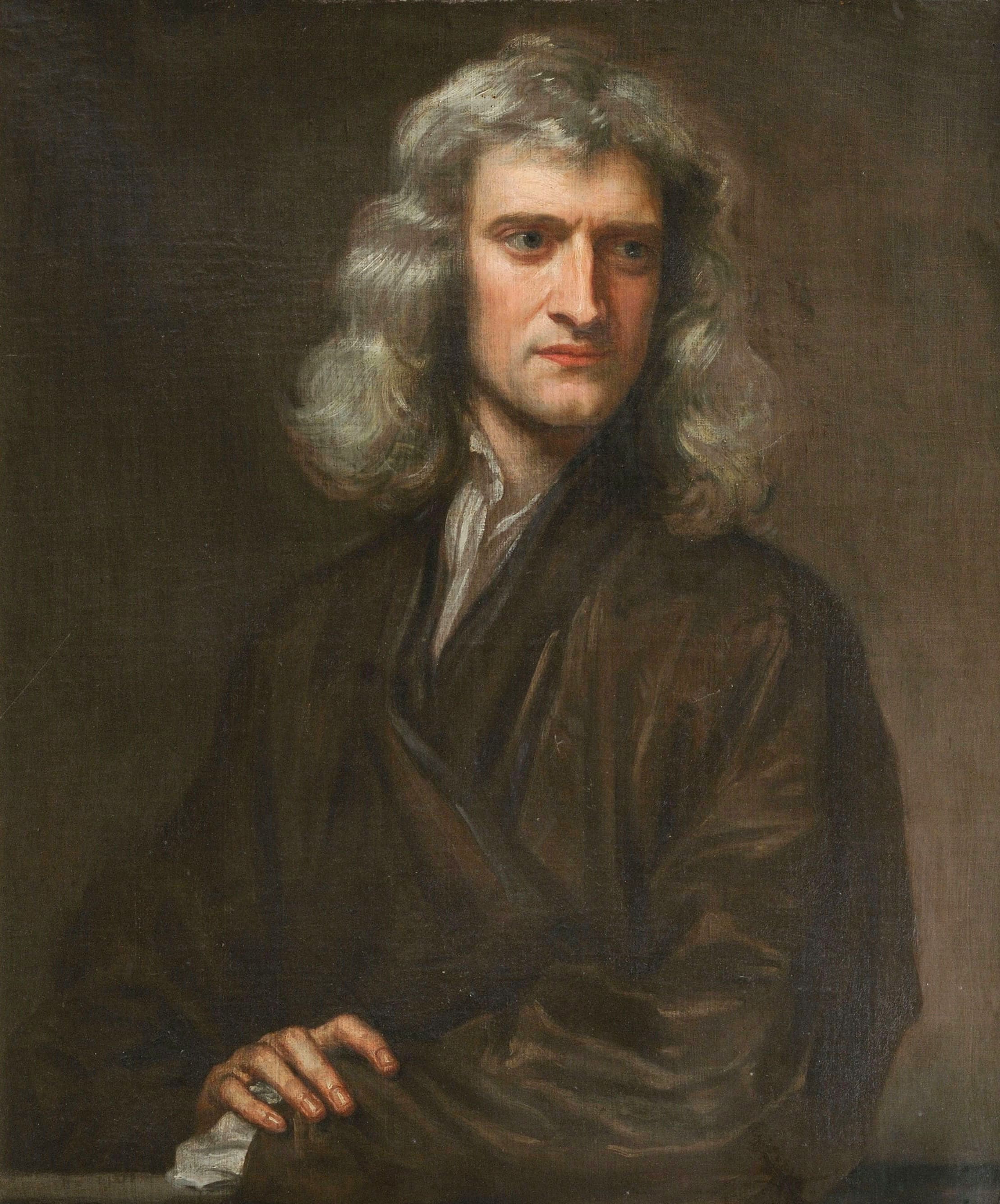}
\end{center}
\caption{\small  Portrait of Isaac Newton by Godfrey Kneller, 1689, oil on canvas, in the private collection of the Earl of Portsmouth at Farleigh House, in Farleigh Wallop, Hampshire.  Image courtesy of the Farleigh Wallop Estate.\la{fig:Newton-Kneller}}
\end{figure*}

On the other hand, a variety of sources allege the existence of an earlier portrait Isaac Newton.  In 1677 mathematician John Collins wrote to Newton that
\begin{quote}
Mr [David] Loggan informs me he hath drawn your effigies in order to a sculpture thereof to be prefixed to a book of Light Colours Dioptricks which you intend to publish, of which we should be glade to have more certain nottice.\cite{Collins1677}
\end{quote}
Newton did not, however, publish any book on optics until 1704 and no trace of Loggan's drawing has been found.

In 1907, Lady Alice Archer Houblon indicated that
\begin{quote}
a portrait, said to be of Sir Isaac as a young man (now at Hallingbury [Place]), was possibly given by him to Sir John [Newton, 3rd baronet of Barrs Court (c.~1651--1734)], or purchased by the latter at the sale of Sir Isaac's effects after his death in 1727.  The baronet survived him
some years, and was represented at the funeral by his son Sir Michael (Knight of the Bath in his father's lifetime), who acted as chief mourner on the occasion.\cite{Houblon}
\end{quote}
Sir John was Isaac Newton's third cousin and the latter regarded the former as head of their common clan.\cite{SirJohn}  Lady Alice, daughter of the 25th Earl of Crawford, was married to Col.\ George Bramston Archer Houblon, who descended from this Sir John Newton.\cite{George}  However, the portrait that Lady Alice reproduces in her book, shown in \Fig{fig:Newton-early}(a), seems to me obviously Continental and therefore not of Isaac Newton.  As far as I can tell, no art historian has repeated the claim that this portrait depicts the famous scientist as a young man.

Milo Keynes records the following strange story about a painting in the the collection of St Catharine's College, Cambridge:
\begin{quote}
In 1669, when he became Lucasian Professor, Newton was aged twenty-six.  A number of pictures, that had been put away behind the College Library, were examined in 1943 or 1944, and one found `much decayed which apparently represented a youngish man suffering from some form of madness.' On restoration and found to be signed by [Henry] Cook and dated 1669, it was now thought to be a portrait of Newton, particularly when the oldest Fellow confirmed the belief that the College had indeed possessed such a portrait.\cite{Keynes-Cook}
\end{quote}

\begin{figure*} [t]
\begin{center}
	\subfigure[]{\includegraphics[height=0.62 \textwidth]{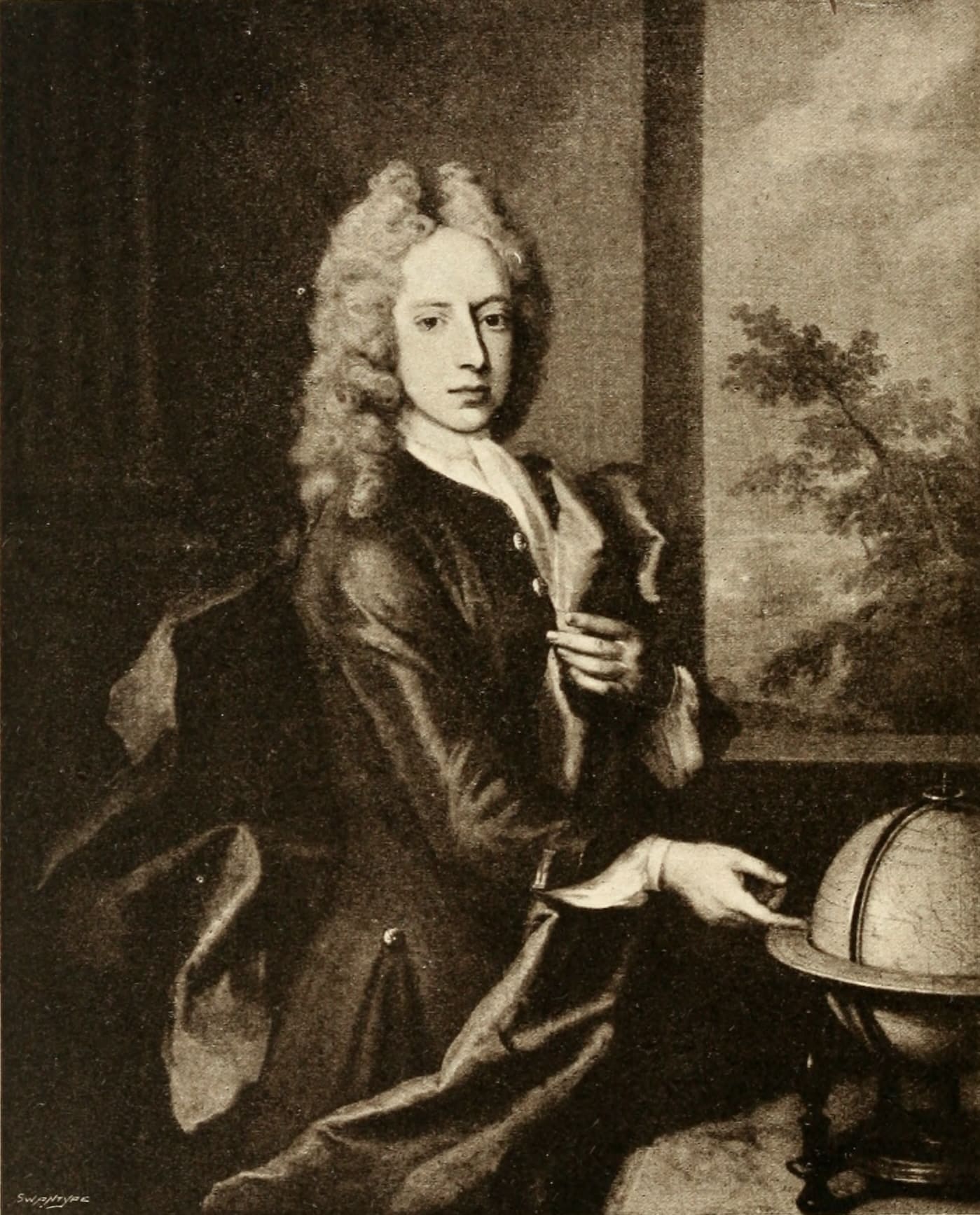}} \hskip 0.3 cm
	\subfigure[]{\includegraphics[height=0.62 \textwidth]{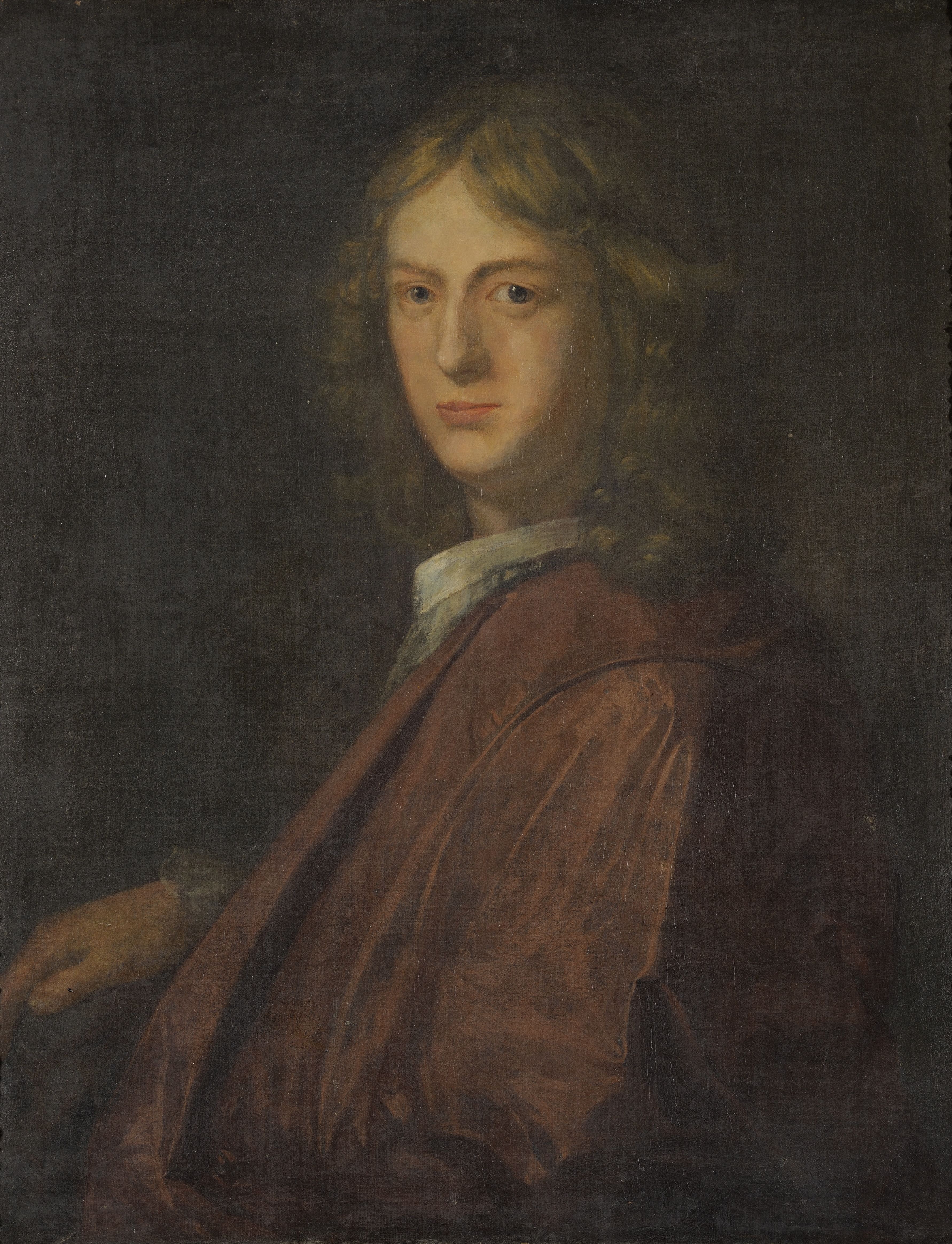}}
\end{center}
\caption{\small (a) Reproduction (Swantype) of a portrait by an unknown artist, from the collection of Col.\ George Bramston Archer Houblon (1843--1913) at Hallingbury Place, near Bishop's Stortford.  This appears in Ref.~\onlinecite{Houblon} with the caption ``Sir Isaac Newton (as a youth)''.  This book is in the public domain.  (b)  Portrait of an unknown young man by Henry Cook, oil on canvas, 1669, collection of St Catharine's College, Cambridge.\cite{Catz-gentleman}  This is identified in Ref.~\onlinecite{Churchill} as ``the earliest portrait of Sir Isaac Newton''.  Image courtesy of the Master and Fellows of St Catharine's College, Cambridge.\la{fig:Newton-early}}
\end{figure*}

\noindent That painting is shown in \Fig{fig:Newton-early}(b).\cite{Catz-gentleman}  It was claimed as a genuine portrait of Newton in a letter to {\it Nature} published in 1945 by the Bursar of St Catharine's, anthropologist John Henry Hutton.\cite{Hutton1}  He repeated that claim in a catalogue of the pictures at St Catherine's, privately printed in 1950.\cite{Hutton2}  A historian of science, Mary S.\ Churchill, publicly declared this as ``the earliest portrait of Sir Isaac Newton'' in 1967.\cite{Churchill}  Keynes, however, noted that it was not considered an authentic picture of Newton by Henry Hake, director of the National Portrait Gallery, a view that has since prevailed.\cite{Keynes-Cook}

Another purported portrait of a young Newton is shown in \Fig{fig:unknown}(a).  This was engraved by Burnet Reading in 1799, ostensibly from a portrait by artist Peter Lely (1618 -- 1680) of Newton as a Bachelor of Arts.\cite{Reading}  It was used as a frontispiece to McGuire and Tamny's 1983 edition of a notebook kept by Newton as a student at Trinity.\cite{notebook} However, no such picture by Lely is known and the sitter bears little resemblance to other portraits of Newton.  Keynes lists this under ``doubtful or spurious portraits''.\cite{Keynes-Reading}  

\begin{figure*} [t]
\begin{center}
	\subfigure[]{\includegraphics[width=0.46 \textwidth]{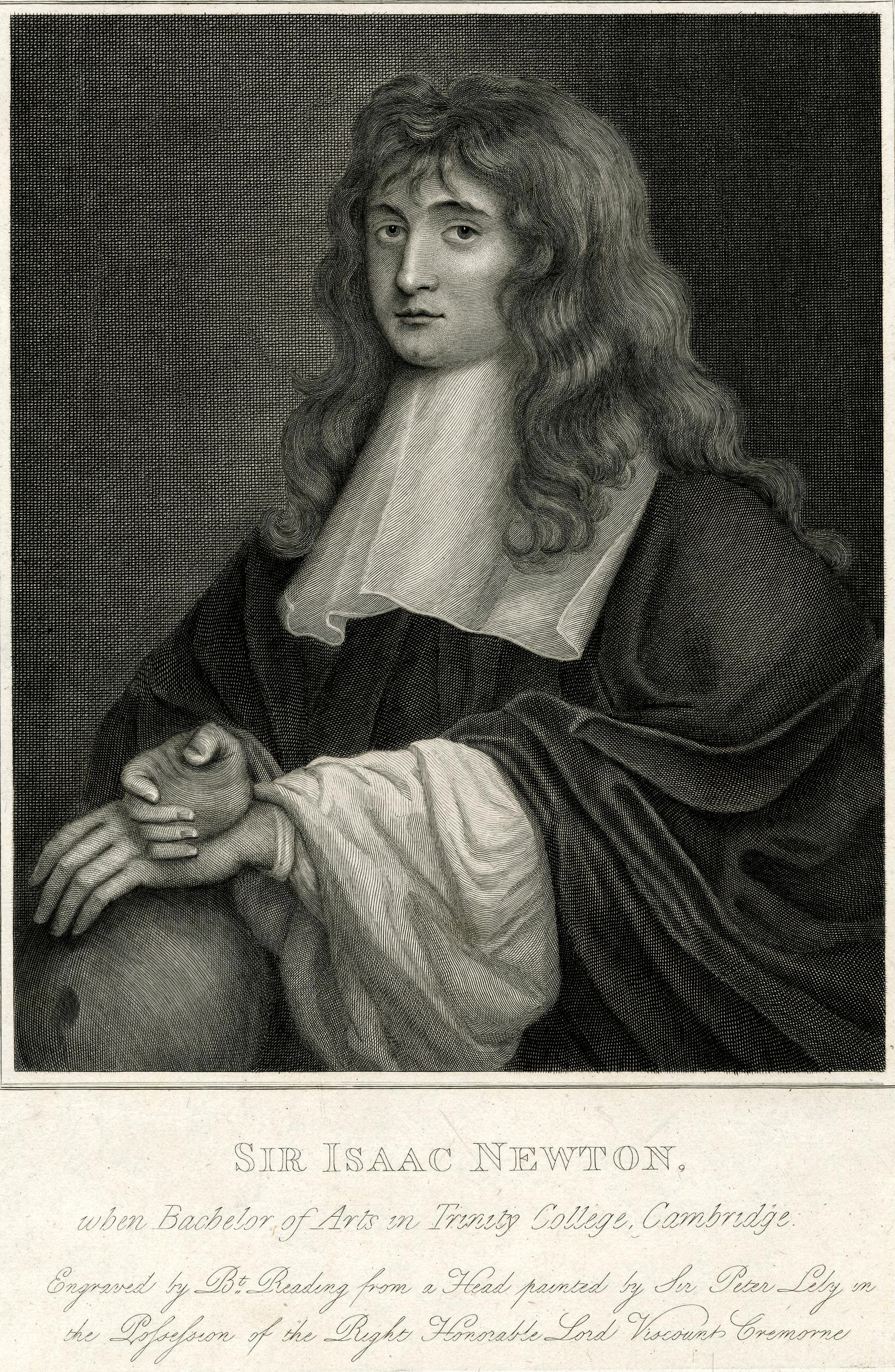}} \hskip 0.25 cm
	\subfigure[]{\includegraphics[width=0.52 \textwidth]{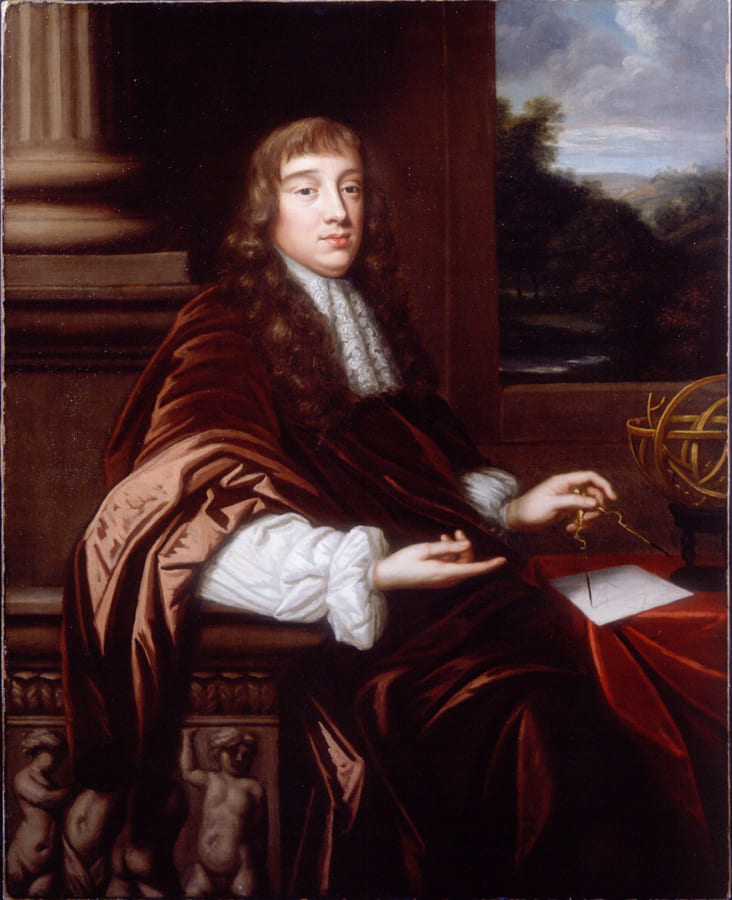}}
\end{center}
\caption{(a) Engraving by Burnet Reading, dated 1799.  The inscription reads {\it Sir Isaac Newton, when Bachelor of Arts in Trinity College, Cambridge.  Engraved by Bt.\ Reading from a Head painted by Sir Peter Lely in the Possession of the Right Honorable Lord Viscount Cremorne.}\cite{Reading}  Image from Apollo - University of Cambridge Repository, reproduced here under the terms of the Creative Commons license.  (b)  Portrait of a mathematician by Mary Beale (1633--1699), oil on canvas, currently in a private collection.\cite{Beale}  Image courtesy of Philip Mould \& Company.\la{fig:unknown}}
\end{figure*}

To the best of my knowledge, neither Keynes nor any other scholar has noted that the young man in Reading's engraving bears a resemblance to the {\it Portrait of a Mathematician} by Mary Beale (1633 -- 1699), shown in \Fig{fig:unknown}(b).  This was incorrectly listed as a portrait of Newton by Kneller when it was auctioned by Christie's in 1964 and 1965.\cite{Beale}  Biologist Lawrence Griffing has argued that it might be the fabled lost portrait of Robert Hooke.\cite{Griffing}  More plausibly, in my opinion, Christopher Whittaker has suggested that it might depict a young Isaac Barrow (1630 -- 1677).\cite{Whittaker}

\begin{figure*} [t]
\begin{center}
	\includegraphics[height=0.62 \textwidth]{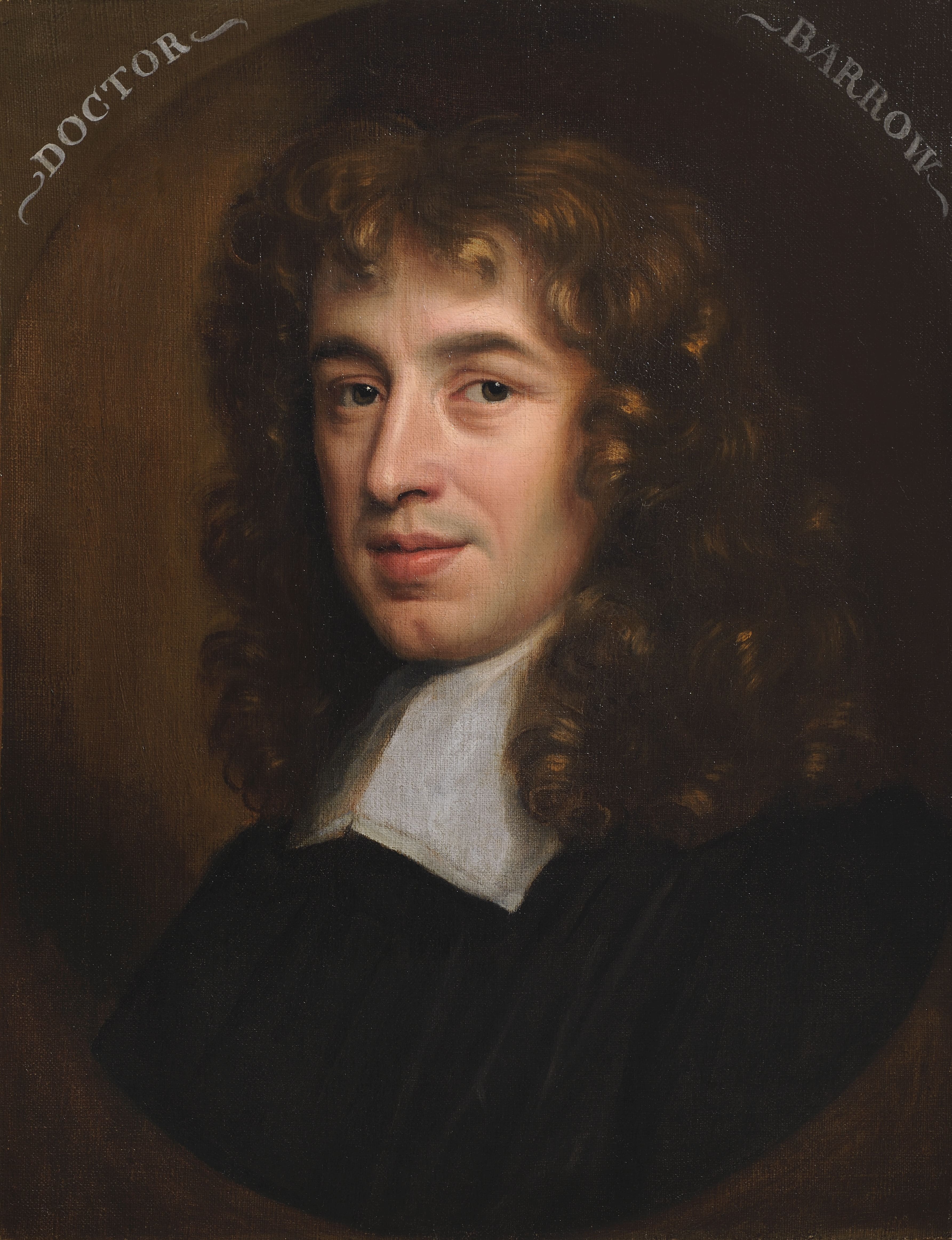}
\end{center}
\caption{Portrait of Isaac Barrow by Mary Beale, oil on canvas, collection of Trinity College, Cambridge (TC Oils P 17).  Image courtesy of the Master and Fellows of Trinity College, Cambridge.\la{fig:Barrow-Beale}}
\end{figure*}

According to his friend and biographer Abraham Hill, Barrow was
\begin{quote}
in person of the lesser size, and lean; of extraordinary strength, of a fair and calm complexion, a thin skin, very sensible of the cold; his eyes grey, clear, and somewhat short-sighted; his hair of a light auburn, very fine and curling.\cite{Hill}
\end{quote}
This is consistent with the \Fig{fig:Barrow-Beale}, which shows Beale's painting of a mature Barrow, now in Trinity's collection.\cite{Barrow-Beale}  Another similar oil painting of Barrow by Beale is in private ownership.\cite{1stDibs}  A drawing of Barrow by artist David Loggan (1634--1692), shown in \Fig{fig:Barrow-Loggan}(a) and dated 1676, also depicts him with a broad chin and curly hair.\cite{Barrow-Loggan}

Hill indicated that Barrow's ``picture was never drawn from life''.\cite{Hill}  This is contradicted by John Ward who wrote that, even though Barrow never sat for a portrait, his picture was painted from drawings that the artist made while Barrow conversed with friends.  A footnote in Ward's {\it Lives of the Gresham Professors} (1740) indicates that the artist was Mary Beale.\cite{Ward}  However, the engraving by Loggan shown in \Fig{fig:Barrow-Loggan}(b), which was used as frontispiece in Barrow's {\it Several Sermons Against Evil-Speaking} (1678) and later, in somewhat modified form, in his collected {\it Works} (1687), carries the inscription {\it ad vivum delin.} (``drawn from life'') and appears to be based on the drawing in \Fig{fig:Barrow-Loggan}(a).\cite{Loggan}

\begin{figure*} [t]
\begin{center}
	\subfigure[]{\includegraphics[height=0.58 \textwidth]{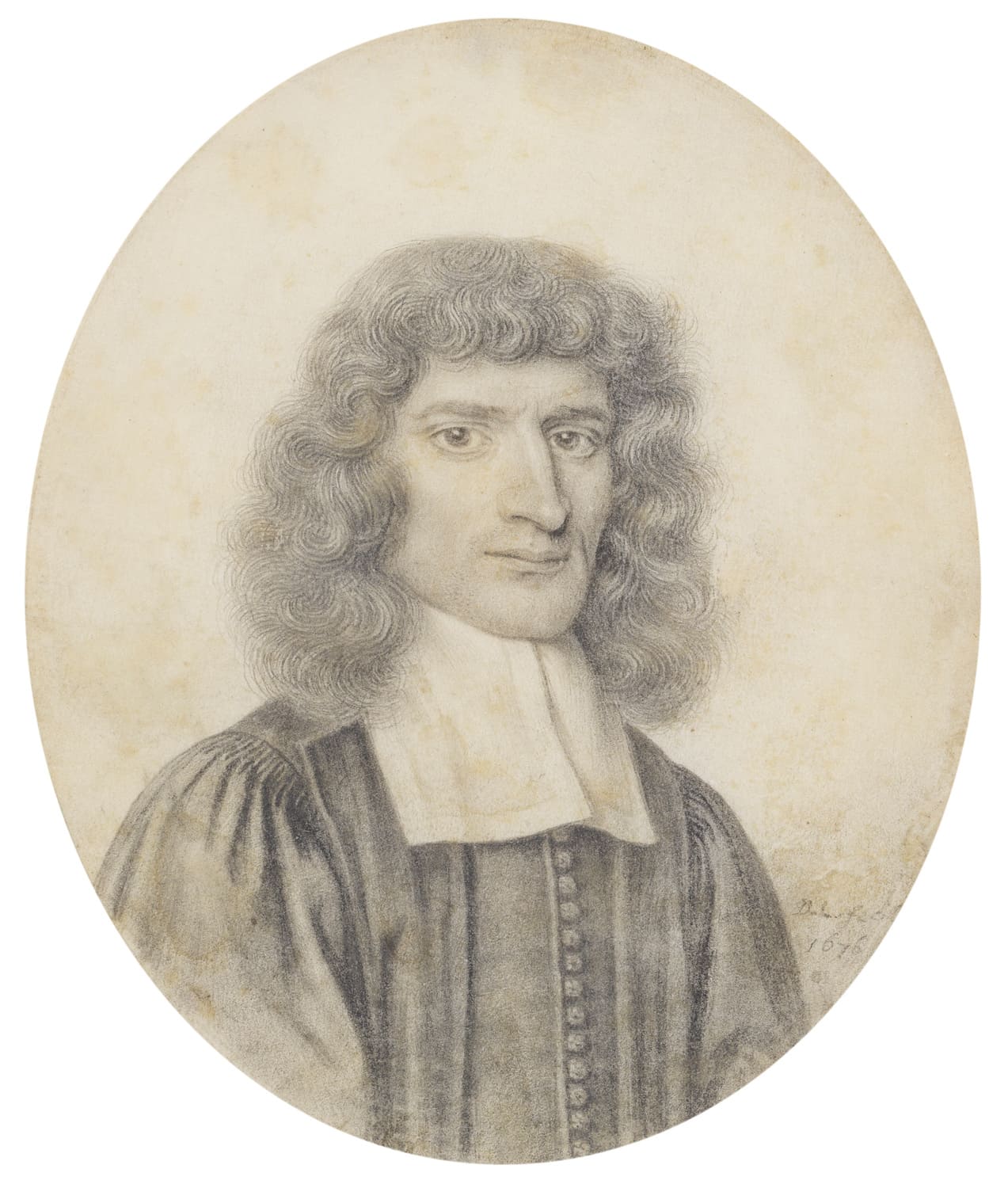}} \hskip 1.5 cm
	\subfigure[]{\includegraphics[height=0.58 \textwidth]{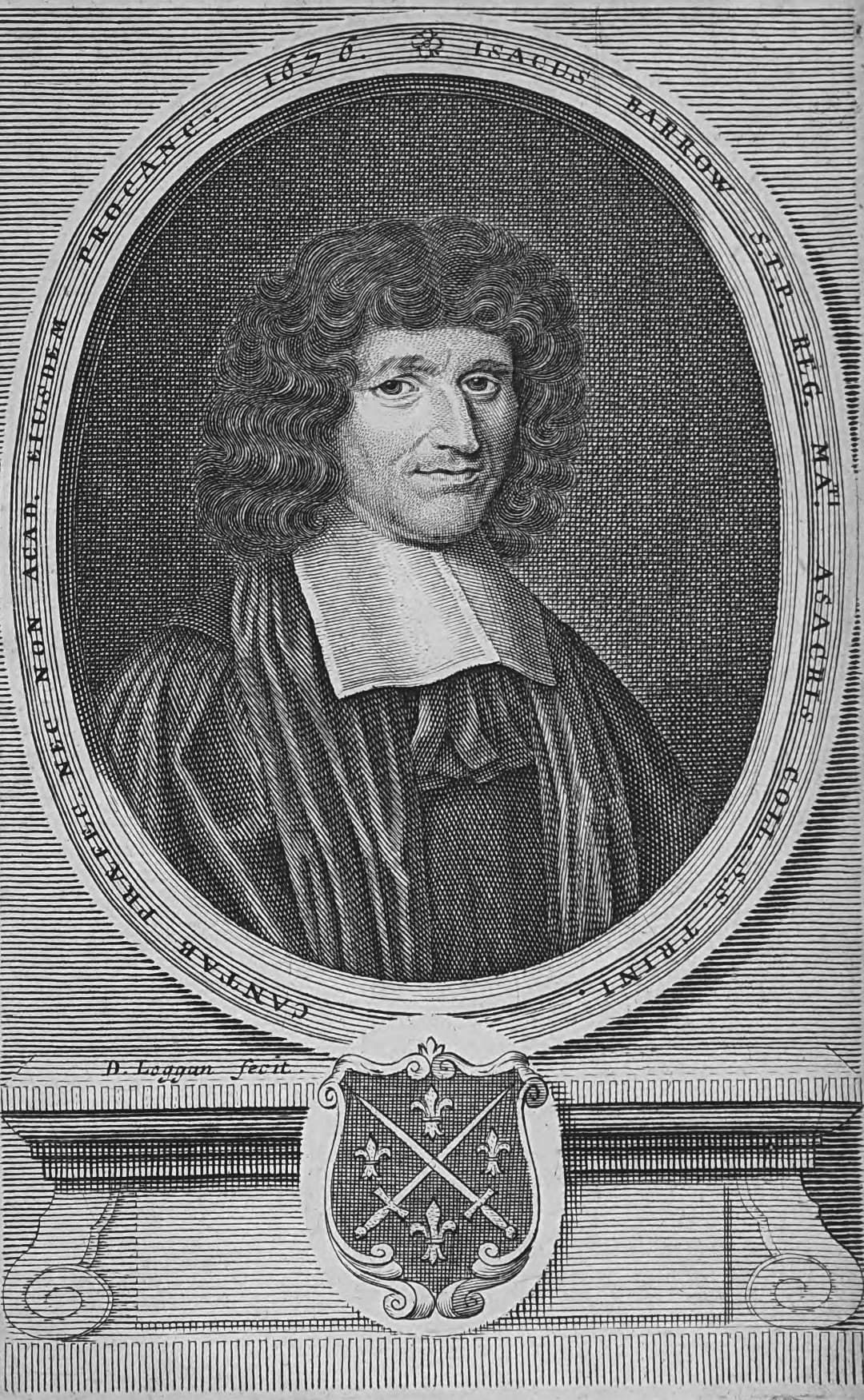}}
\end{center}
\caption{(a)  Drawing of Isaac Barrow by David Loggan, plumbago on vellum, 1676, National Portrait Gallery, London (NPG 1876). (b)  Engraving by Loggan used as frontispiece to Barrow's {\it Several Sermons Against Evil-Speaking} (1678).  Images courtesy of the National Portrait Gallery, London.\la{fig:Barrow-Loggan}}
\end{figure*}

In my opinion, the portrait of \Fig{fig:unknown}(b) is consistent with the physical features of Barrow in \Fig{fig:Barrow-Beale}, although evidently the older man is much reduced in bulk.  As quoted above, Hill described Barrow as short, lean, and very strong.  He also recounts how Barrow helped fight off an attack by Barbary pirates during his passage to Constantinople in 1657.\cite{Hill}  It is possible, however, that Barrow might have been less athletic in earlier years.  If art expert Tabitha Barber is correct in dating Beale's {\it Portrait of a mathematician} to c.\ 1680,\cite{Beale} and if it is indeed a portrait of Barrow, then it must be based on some other depiction of Barrow made decades earlier, perhaps connected to the unknown source of Reading's engraving in \Fig{fig:unknown}(a).  In any case, here we will be mainly concerned with a different puzzle than the identification of the stout young scholar of \Fig{fig:unknown}, who is certainly not Isaac Newton.

%%%%%%%%%%
%%% THE TAYLER PICTURE
%%%%%%%%%%
\section{The Tayler picture}
\la{sec:Tayler}

\begin{figure*} [t]
\begin{center}
	\includegraphics[width=0.58 \textwidth]{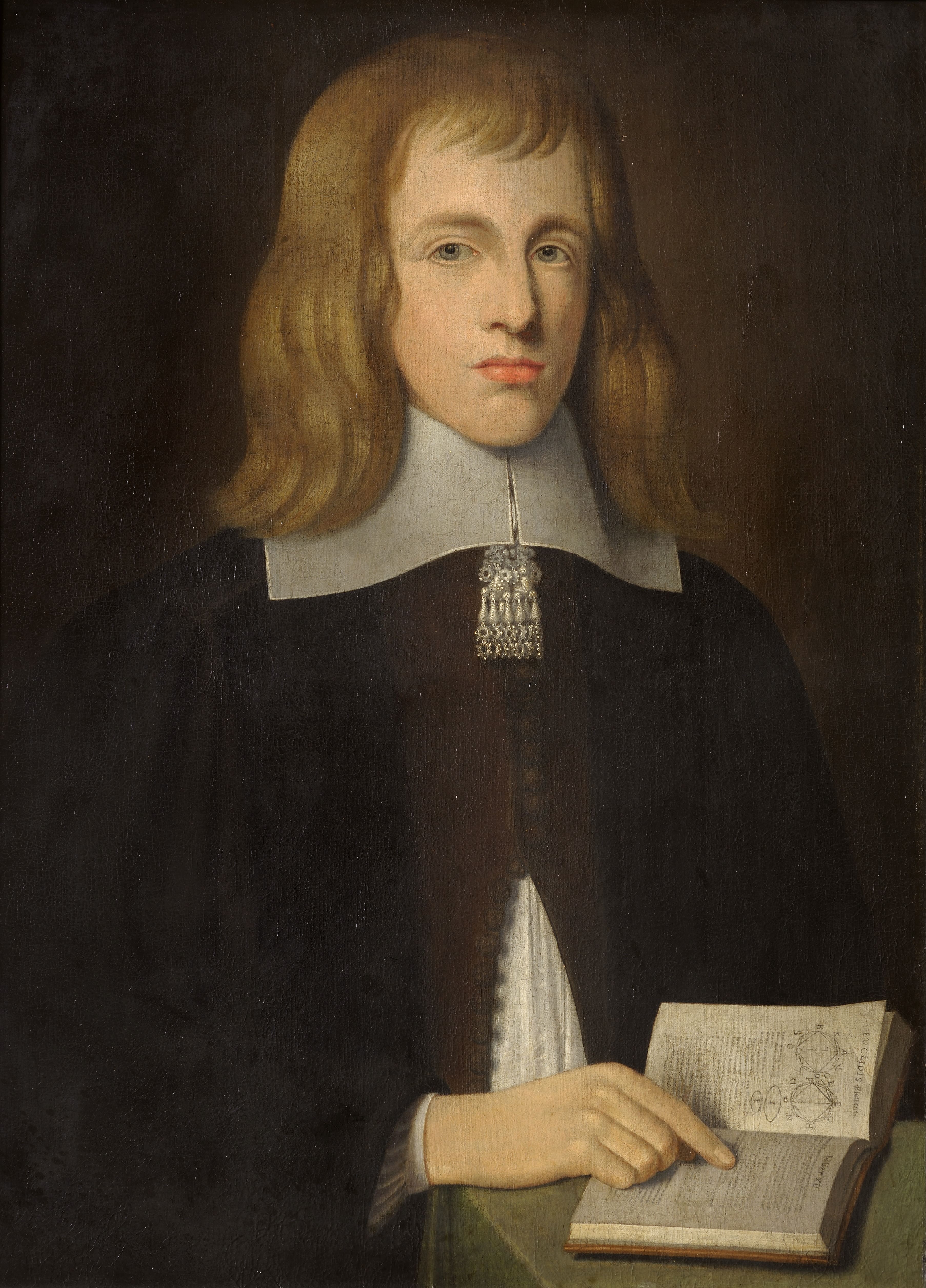}
\end{center}
\caption{\small ``Tayler picture'': unknown painter, oil on canvas, collection of Trinity College, Cambridge (TC Oils P 16).  It currently hangs in a private room of the Master's Lodge.  It was formerly owned by the Rev.\ G.~W.~H.~Tayler (1831--1890) and, before him, by his uncle the Rev.\  C.~B.~Tayler (1797--1875), both of whom were graduates of Trinity and personal friends of the physicist James Clerk Maxwell.  Image courtesy of the Master and Fellows of Trinity College, Cambridge.\la{fig:Tayler}}
\end{figure*}

Figure \ref{fig:Tayler} shows a portrait in oils that currently hangs in a private room in the Master's Lodge of Trinity College, Cambridge.\cite{Tayler}  This work by an unknown artist shows a young blond and blue-eyed man, dressed in academic gown and bands.  The bands are very wide, and under them the collar is tied with white tassels of an unusual, ornate design.  The dark gown has bell-shaped sleeves and is open at the front.  Although this is difficult to ascertain in the picture's current state of conservation, the gown appears to be decorated with facings of a different material, perhaps velvet.  The lower buttons on the coat are undone, showing the folds of the shirt underneath.  The most salient aspect of the portrait is that the sitter's right hand points to a passage in an open mathematical book.  The artist has reproduced the pages in the book with sufficient detail that they can be unequivocally identified as pages 292 and 293 in Isaac Barrow's 1655 Latin edition of Euclid's {\it Elements}.  These correspond to book 12, proposition 2: ``Circles are to one another as the squares of their diameters''.

According to documents provided by Nicolas Bell, the librarian of Trinity College, this painting was purchased by the college in 1924 from George Herbert Last (1877--1949) of 25 The Broadway, Bromley.  Last was a primarily a dealer in rare books and manuscripts, not in art.\cite{Last}  He had acquired the picture from the daughter of Rev.\ George Wood Henry Tayler (1831--1890), vicar of Holy Trinity Church, Carlisle, who had inherited it from his uncle the Rev.\ Charles Benjamin Tayler (1797--1875), rector of St Peter's Church, Chester and of Ottley, Suffolk.\cite{CBTayler}  In response to an inquiry from mechanical engineer David Pye, who was a Fellow of Trinity, Last explained that
\begin{quote}
Both of these gentlemen were Master of Arts of Trinity College, Cambridge.  How [the picture] came into the possession of the Tayler family is quite unknown, and neither of the former owners could say definitely whose portrait is was, but it was said among them, it was that of Isaac Barrow.\cite{Last-Pye}
\end{quote}
The Taylers had close ties not only to Trinity, but also to the great theoretical physicist James Clerk Maxwell (1831--1879).\cite{Maxwell}  Whether this had something to do with their acquisition of the portrait is unknown.  Henceforth I will refer to the painting shown in \Fig{fig:Tayler} as the {\it Tayler picture}.

A brass plaque attached to the frame now identifies the Tayler picture as depicting Isaac Barrow, who was Master of Trinity from 1672 until his untimely death in 1677, aged only 46.  Milo Keynes included it as a portrait of Barrow in an article on ``The personality of Isaac Newton'' published in 1995.\cite{Keynes-personality}  However, I see little resemblance with the Barrow of Figs.\ \ref{fig:Barrow-Beale} and \ref{fig:Barrow-Loggan}.  When Trinity purchased the Tayler picture in 1924, it was thought that it might be the work by Beale mentioned in John Ward's biography of Barrow.\cite{Winstanley}  The subsequent emergence of the portrait reproduced in \Fig{fig:Barrow-Beale} makes this untenable.

In my opinion, a more likely candidate for the sitter of the Tayler picture is Isaac Newton.  If there is indeed a lost portrait of a young Newton, as a longstanding tradition alleges, then the Tayler picture seems to me the best candidate available.  Did a confusion between Trinity's two great mathematical Isaacs ---who succeeded each other as Lucasian Professor--- lie behind more than one misidentification of a 17th-century portrait?

\begin{figure*} [t]
\begin{center}
	\includegraphics[height=0.75 \textwidth]{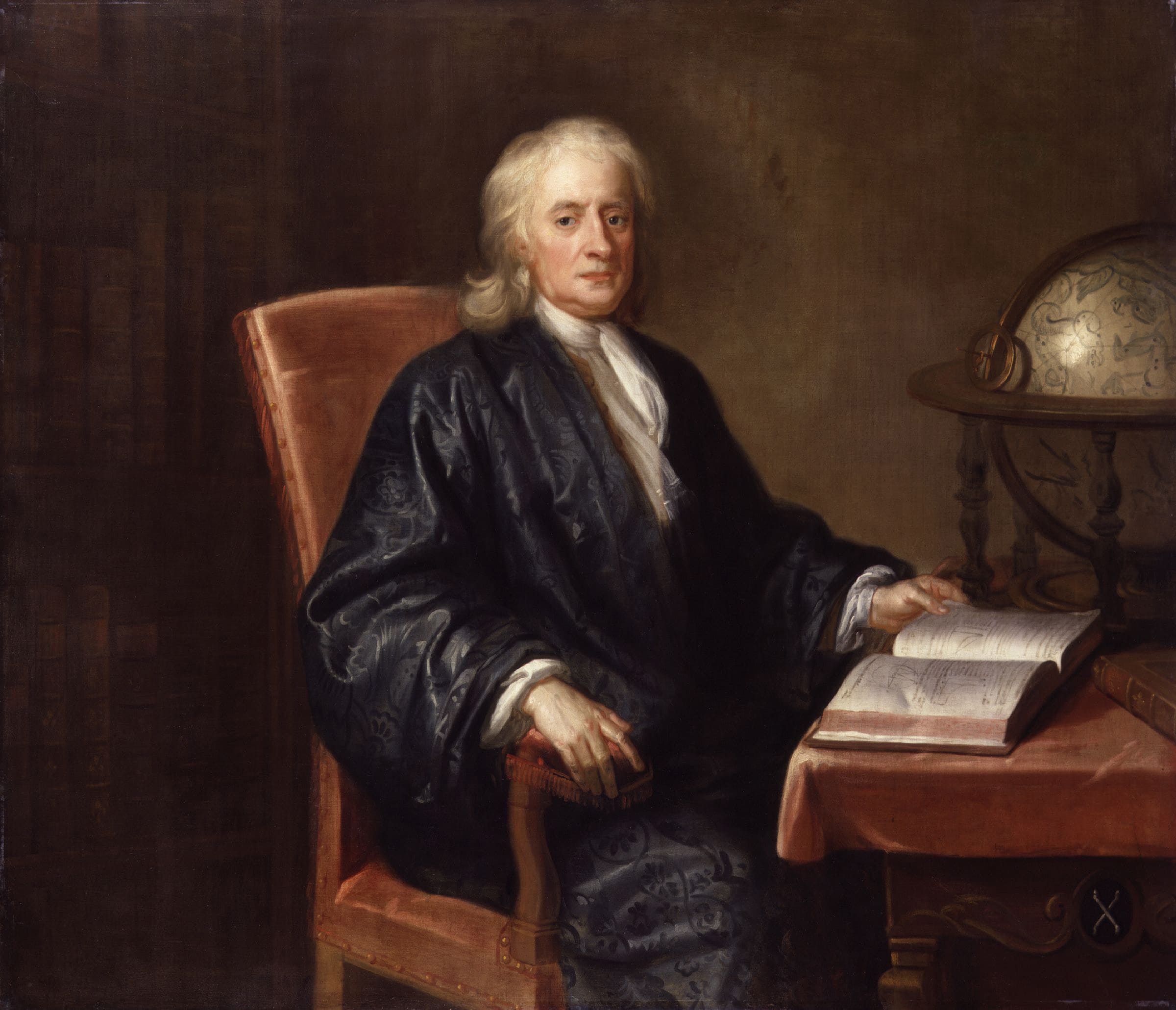}
\end{center}
\caption{\small Portrait of Isaac Newton by the studio of Enoch Seeman, c.\ 1726, oil on canvas, National Portrait Gallery, London (NPG 558).  Image courtesy of the National Portrait Gallery.  \cite{Seeman-NPG}\la{fig:Seeman-NPG}}
\end{figure*}

In the better of the genuine {\it ad vivum} portraits, Newton is depicted as blue-eyed,\cite{Keynes-blue} with an exceptionally high forehead, a thin and prominent nose, full lips, and (when not shown wearing a wig) a full head of straight hair.  According to Richard Atterbury, the Bishop of Rochester and Jacobite conspirator,
\begin{quote}
in the whole of his face and make, there was nothing of that penetrating sagacity which appears in his composures.  He had something rather languid in his look and manner, which did not raise any great expectation in those who did not know him.\cite{Atterbury}
\end{quote}
Philosopher Henry More, an older contemporary of Newton at Cambridge, wrote that Newton's ordinary countenance was ``melancholy and thoughtful''.\cite{More}

Figure \ref{fig:Seeman-NPG} shows a portrait of Newton seated before a table with a celestial globe and a copy of the third edition of his {\it Principia} (1726), open to book I, section 12, proposition 81 (pp.\ 204-5).\cite{Seeman-Chandra}  The unfinished background shows the ghostly outline of a bookshelf.  This is a copy of a portrait of the elderly Newton by Enoch Seeman the younger (c.\ 1689 -- 1745), now in a private collection, which has a very similar but somewhat simpler composition.\cite{Seeman-Christies}  According to Milo Keynes, the use of the heraldic decoration (a pair of crossbones) on the side of the table suggests that the copy may have been made for the Newton family by Seeman's workshop.\cite{Seeman-Keynes}  The similarities between this and the Tayler picture strike me as tantalizing.  The depiction of Newton's face in \Fig{fig:Seeman-NPG} closely follows an earlier portrait by Seeman, dated 1726,\cite{Seeman-TC} which is shown in \Fig{fig:Seeman-TC}.

\begin{figure*} [t]
\begin{center}
	\includegraphics[height=0.62 \textwidth]{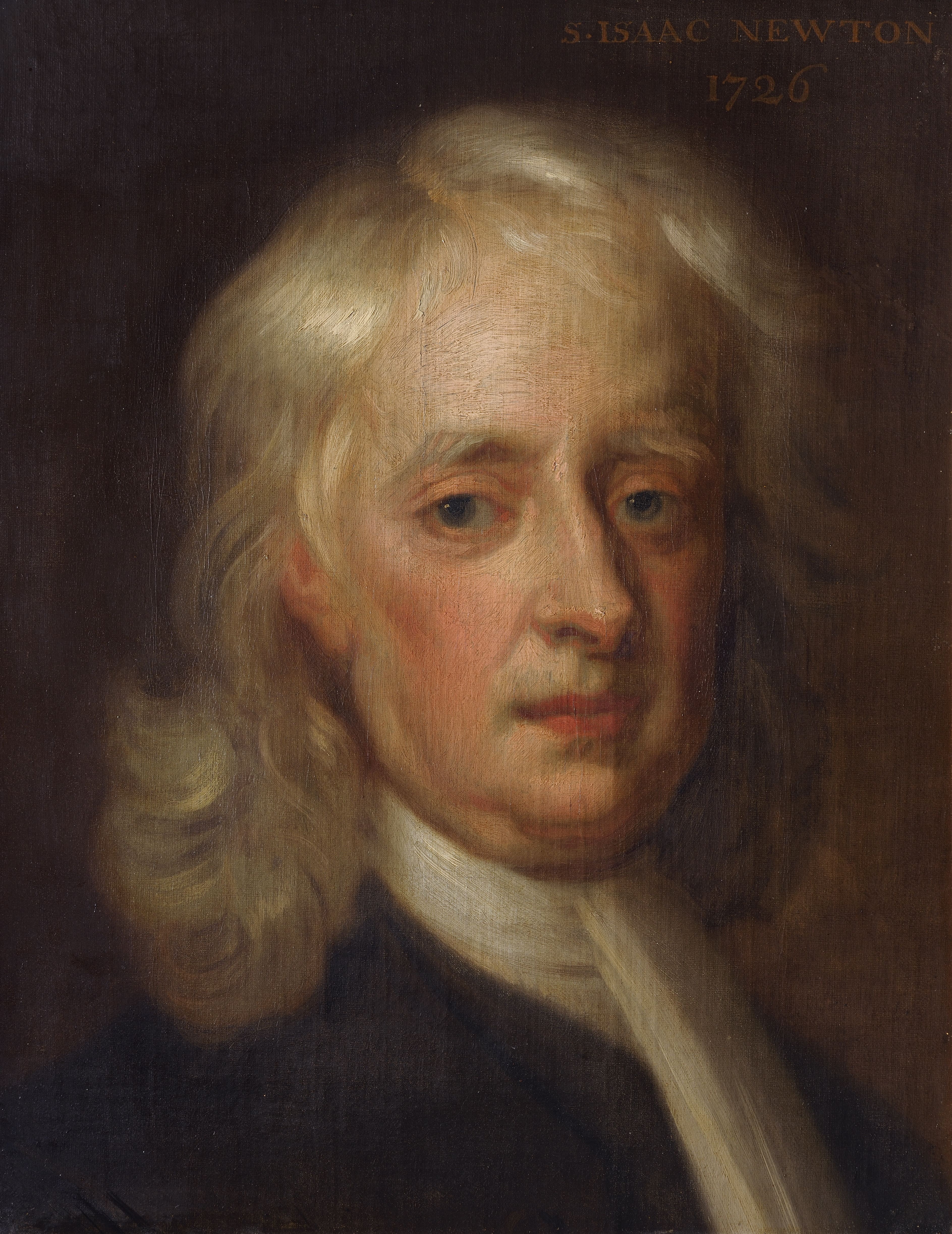}
\end{center}
\caption{\small Portrait of Isaac Newton by Enoch Seeman the younger, 1726, oil on canvas, collection of Trinity College, Cambridge (TC Oils P 138).\cite{Seeman-TC}  Image courtesy of the Master and Fellows of Trinity College, Cambridge.\la{fig:Seeman-TC}}
\end{figure*}

In all the known portraits of a wigless Newton, his hair is shown as grey or white.  Both John Wickins (his roommate at Cambridge from 1665 to 1683) and Humphrey Newton (who was not a relation and who served as his secretary from 1685 to 1690, in the years surrounding the composition of the {\it Principia}) reported that Isaac Newton's hair was already gray at the age of thirty.\cite{NaR-hair}  Wickins recalled that Newton jokingly attributed the fact that his hair had turned grey so soon to his frequent use of mercury in alchemical experiments.\cite{Wickins}  The importance attached to mercury by traditional alchemists was due in part to its ability to dissolve (``amalgamate'') most other metals, including gold.  If Newton had previously been blond, that joke might refer to the amalgamation of gold, a process that he must have often witnessed.

In 1728, Newton's friend and collaborator Nicolas Fatio de Duillier published an eclogue in Latin to mark the first anniversary of Newton's death.  That poem contains a mythological allegory reminiscent of a genre of Renaissance alchemical literature:  The Olympian gods pay homage to the newborn Newton.  Venus places around his head a wreath made from Apollo's laurel crown and flowers plucked from her own bosom.  Later, this wreath is stolen by Mercury, whereupon Apollo replaces it with a garland of white flowers and prophesies that ``before your hair has turned white your age shall name you with worship''.  Since Apollo was traditionally represented as golden-haired ({\it chrysocomes}), it seems likely that Fatio intended to indicate that the young Newton had been blond.  Fatio states only, in his preamble to his poem, that Newton ``reached the prime of his life as the colour of his hair gradually changed to perfect white between the age of thirty and forty''.\cite{Fatio}  Fatio first arrived in England in 1687, when Newton was well into his forties.  He would eventually become Newton's principal alchemical collaborator in the period between 1689 and 1694.\cite{alchemy}  Fatio probably intended his allegory to connect the change of Newton's hair with the amalgamation of gold by mercury, which was an important process in the alchemy that he and Newton had practiced together, and which they believed to be connected with the preparation of the philosopher's stone.\cite{sophic}  It is, therefore, conceivable that Fatio might have had in mind a portrait of a fair-headed young Newton ---whom he had not known in person--- when he composed his eclogue to mark Newton's passing.

%%%%%%%%%%
%%% EUCLID, BARROW & NEWTON
%%%%%%%%%%
\section{Euclid, Barrow \& Newton}
\la{sec:Euclid}

Aside from the physical dissimilarities between the Tayler picture and other depictions of Barrow, the details of Barrow's early career make it unlikely that he would have posed with his first edition of Euclid.  Figure \ref{fig:Euclid}(a) shows the book pages from the portrait in question, in closeup and displayed upright.  The indentation of the text below the diagrams is consistent with the first edition of Barrow's translation of Euclid, dated 1655, and not with the subsequent editions of 1657 or 1660.   Far from expressing pride in that first edition, at the time Barrow described it to friends as ``but a meane worke'' that he had composed in haste and which would require corrections.\cite{Feingold-meane}  When publisher William Nealand printed that book on February 1656 (still reckoned as 1655 in the ``old style'' in which New Year's Day was 25 March), Barrow was no longer in England.  He had departed for the Continent eight months earlier, benefiting from a travel grant that offered respite from the Royalist Barrow's clashes with the authorities of Puritan Cambridge.  Barrow returned to England only in September 1659, after the death of Oliver Cromwell and with the restoration of the monarchy imminent.\cite{Feingold-dates}

\begin{figure*} [t]
\begin{center}
	\subfigure[]{\includegraphics[height=0.5 \textwidth]{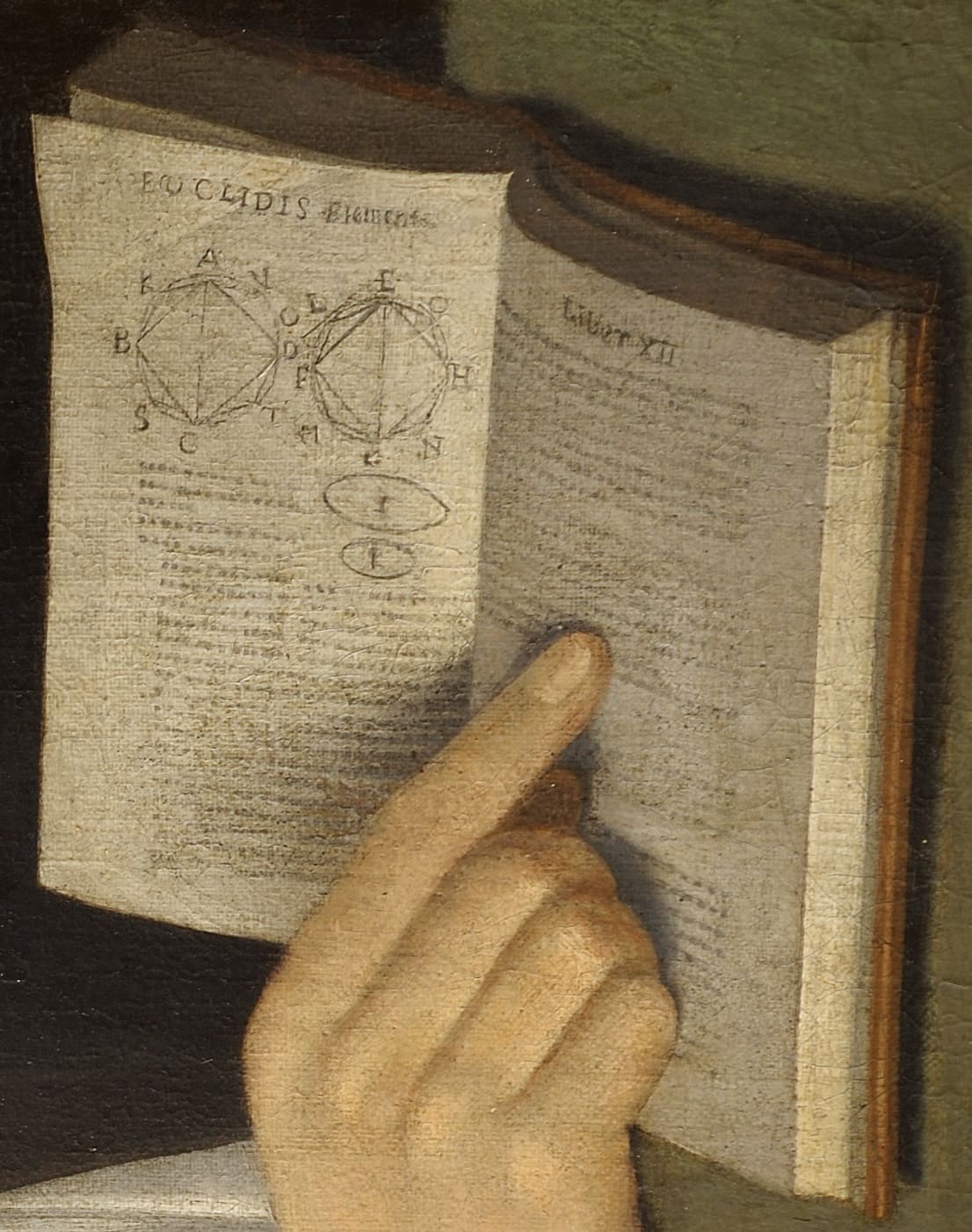}} \hskip 0.4 cm
	\subfigure[]{\includegraphics[height=0.47 \textwidth]{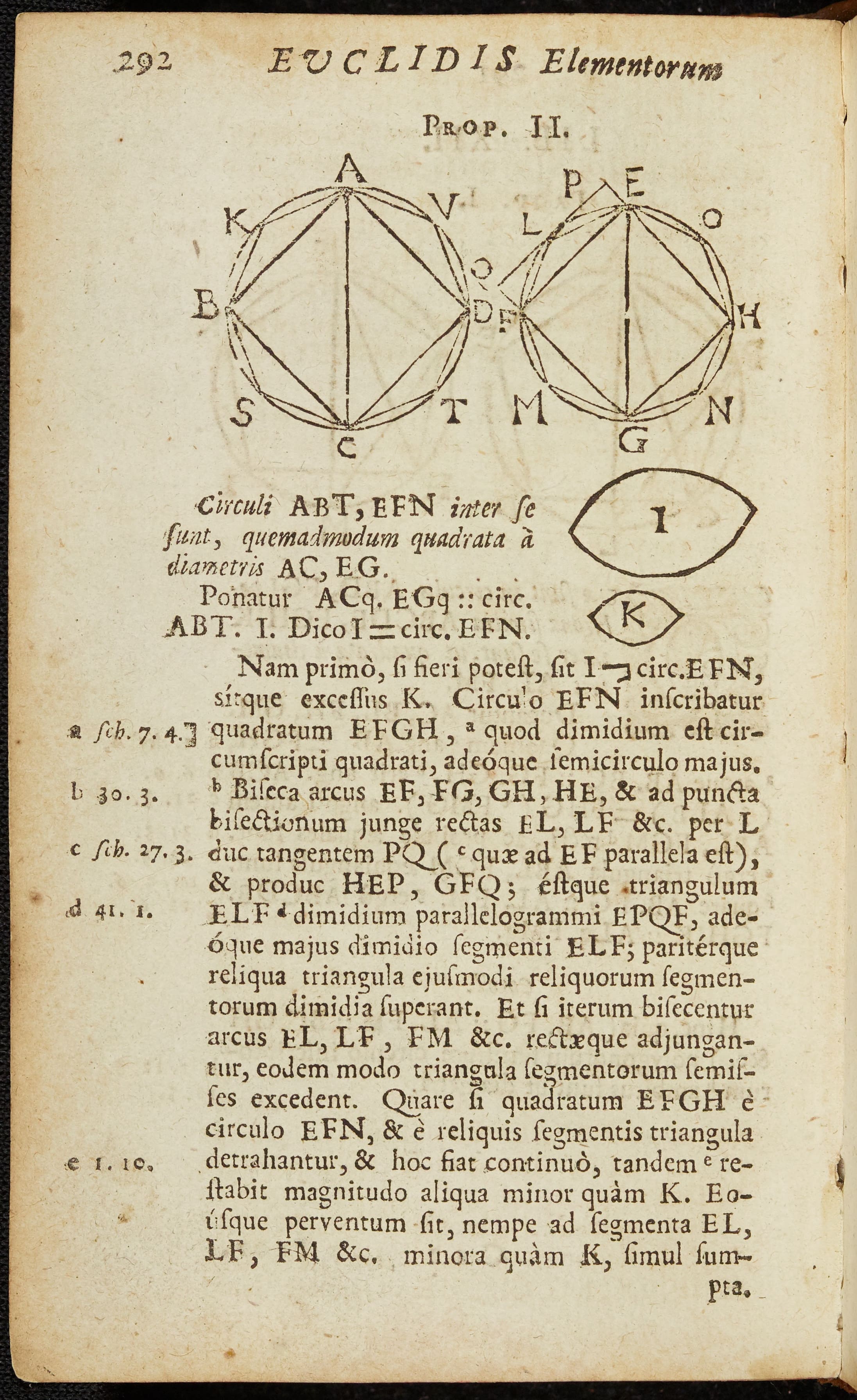}}
	\subfigure[]{\includegraphics[height=0.47 \textwidth]{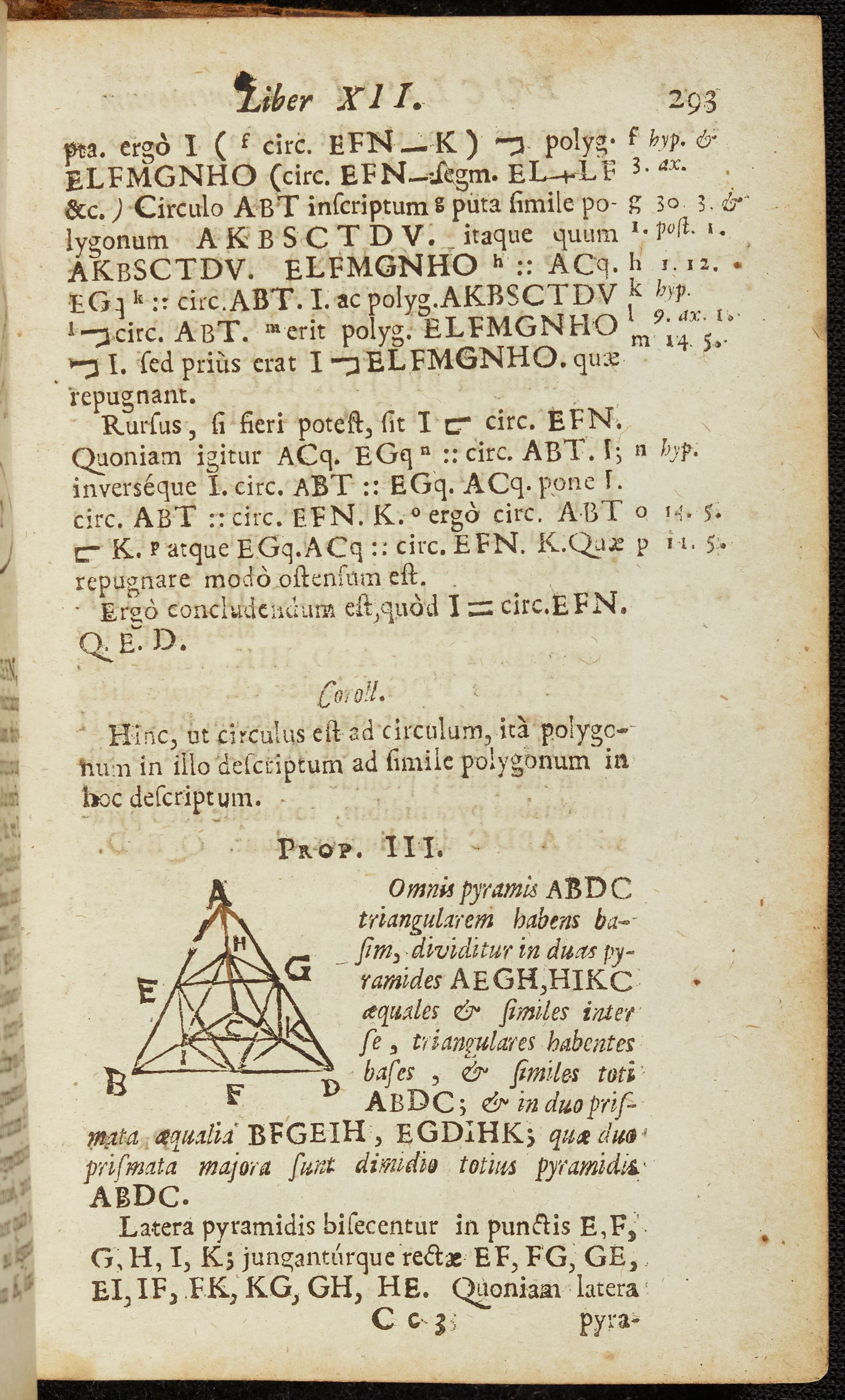}}
\end{center}
\caption{\small (a) Detail from the Tayler picture of \Fig{fig:Tayler}, showing the pages from the book, displayed upright.  (b) Page 292 in Newton's personal copy of Barrow's 1655 edition of Euclid's {\it Elements}, now in the Trinity College library (H581).  (c) Page 293 in the same copy.  Someone (presumably Newton) has filled out in ink the missing line segments at the top of the printed diagram for pyramid $ABCD$.  Images courtesy of the Master and Fellows of Trinity College, Cambridge.\la{fig:Euclid}}
\end{figure*}

Who, if not its author, would pay to have his portrait painted with that book?  In later life, Newton related that, as a student, he initially neglected the study of Euclid in favor of the modern analytical geometry of Descartes.  Sent by his tutor Benjamin Pulleyn to be interviewed by Barrow, Newton made a poor first impression due to his ignorance of Euclid, which Newton was then at pains to remedy.\cite{NaR-Euclid}  The fact that the portrait of \Fig{fig:Tayler} shows the sitter pointing to a proposition in Barrow's Latin edition of Euclid that treats the area of a circle might be consistent with Newton either courting or celebrating Barrow's support of his advancement at Cambridge. 

Newton's own copy of that first edition (item 581 in Harrison's catalog of Newton's books) is now in the Trinity College library.\cite{H581} At some time in the early 1660s, Newton acquired that book and annotated it extensively, correcting errors in Barrow's marginal references, drawing auxiliary lines on diagrams, translating verbal propositions into algebraic notation, and occasionally attempting his own alternative proof of a given result.\cite{Whiteside}  Newton's annotations to this specific tome have recently attracted the interest of historian of mathematics Benjamin Wardhaugh, who dedicated to them a chapter in his {\it Encounters with Euclid}.\cite{Wardhaugh}

Note in \Fig{fig:Euclid}(a) that the artist depicted one of the pages with a folded corner.  This is reminiscent of Newton's habit of dog-earing his books to mark passages of interest,\cite{dog-ear} although no fold is in evidence at that place in \Fig{fig:Euclid}(b) (as I have corroborated by examining Newton's copy in the Trinity College library).  It would appear, in any case, that the artist wished to underline that the sitter had studied the book closely.  In the passage depicted in \Fig{fig:Euclid}, Euclid applies the ``method of exhaustion'' that he had introduced in Book 10, proposition 1.  Newton referred explicitly to Euclid's method of exhaustion at the very end of his {\it De analysi}, composed in 1669, in which he developed the theory of power series.\cite{Deanalysi}

Newton probably attended Barrow's lectures on geometry, which in their published form of 1670 contain a geometrical proof of the fundamental theorem of calculus, based on an exhaustion method similar to the one invoked in the passage of Euclid represented in \Fig{fig:Euclid}.\cite{calculus}  The great influence of Euclid upon the style of the mature Newton's {\it Principia} has often been noted.  Mastering Euclid's {\it Elements} was the first of the recommendations that Newton gave in 1691 to Richard Bentley, the leading classicist of the day and redoubtable Master of Trinity College, when the latter enquired about how to prepare for personal study of the {\it Principia}.\cite{Bentley}  Mathematician Charles R.\ Leedham-Green, who has published the most recent full English translation of the {\it Principia}, notes in his preface that ``Newton regards Euclid as the gold standard.  Thus Newton believes that he can (or that he should) reduce all his mathematics to Euclidean geometry.''\cite{Principia-LG}

Recently, the intellectual historian and Newton scholar Dmitri Levitin has emphasized the great direct influence that Barrow's ``mixed-mathematical'' intellectual program had upon the development of Newton's anti-metaphysical approach to natural philosophy.  In this context, Levitin even quotes from Barrow's comments on Euclid's use of the ``method of superposition'' to demonstrate the congruence of triangles in book 1, proposition 10 of the {\it Elements}.\cite{mixed}

It is not altogether unusual for portraits of that period to show a young scholar studying an identifiable textbook by a more senior author.  Consider Isaac Fuller's painting of William Petty (1623--1687) shown in \Fig{fig:Petty}.\cite{Petty-portrait}  This shows Petty dressed as a Doctor of Medicine of the University of Oxford (a title that he received in 1649), holding a human skull in his left hand and pointing with his right hand to an anatomical illustration in Adriaan van den Spiegel's {\it De Humani Corporis Fabrica} (``On the Fabric of the Human Body'', 1627, plate 3).\cite{Piper-Petty}  This was probably intended to advertise that Petty had studied medicine in the Low Countries and France, before moving to Oxford in 1646.  Petty later became one of the founder Fellows of the Royal Society in 1660, was knighted in 1661, and is today remembered chiefly as a pioneering economist.\cite{Petty}

Another reason why it may be useful to consider here Petty's portrait is that John Aubrey's {\it Brief Lives} describes Petty as having eyes ``a kind of goose-grey, but very short sighted''.\cite{Piper-Petty}  Compare this to Hill's description, quoted above, of Barrow's ``eyes grey, clear, and somewhat short-sighted''.\cite{Hill} Upon close examination, I find that \Fig{fig:Barrow-Beale} and \Fig{fig:Petty} show their respective sitters with what we would today describe as grey eyes, distinguishable from the clear blue eyes of \Fig{fig:Tayler}.  This provides further support for the conclusion that the Tayler picture does not depict Isaac Barrow.

\begin{figure*} [t]
\begin{center}
	\includegraphics[height=0.7 \textwidth]{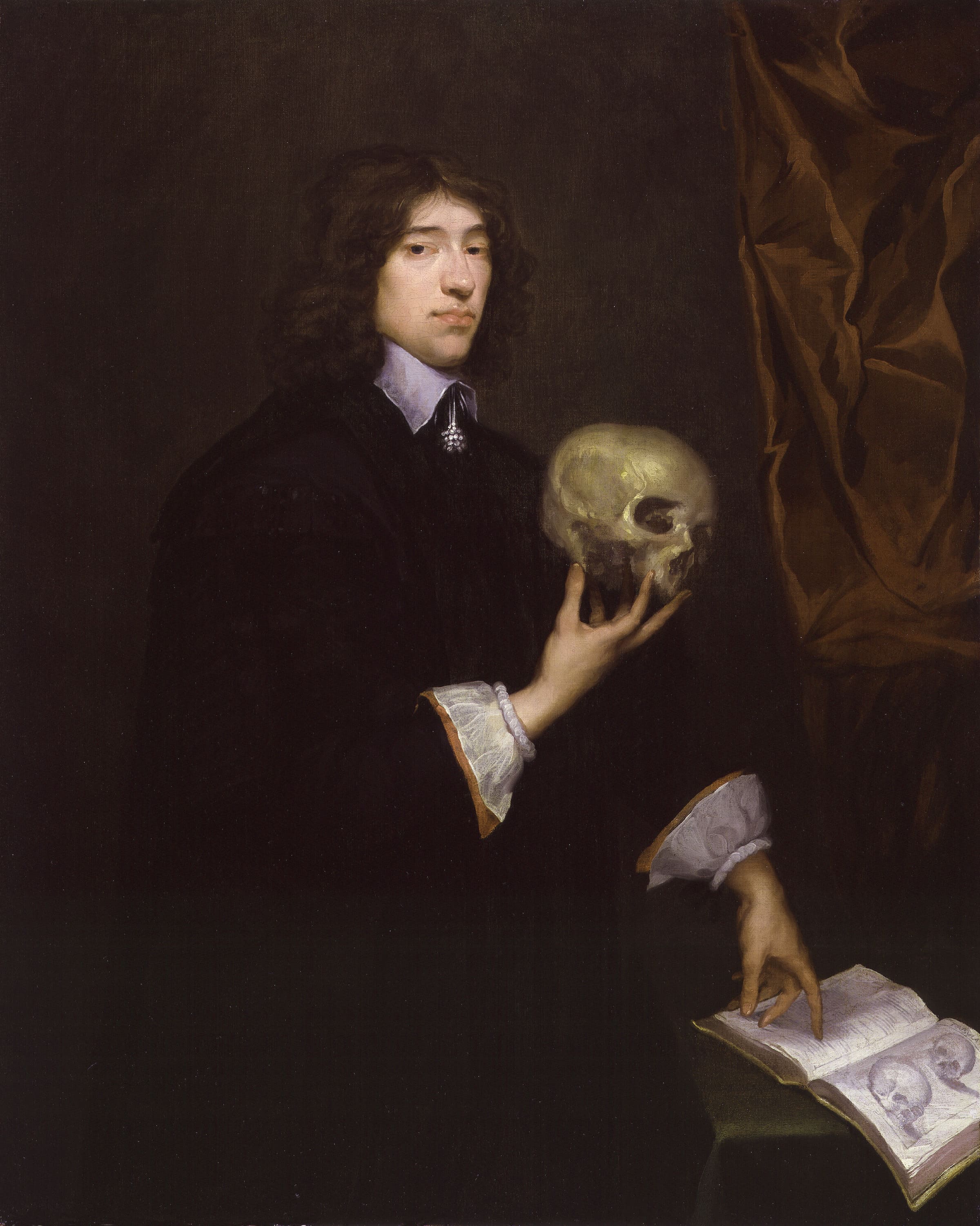}
\end{center}
\caption{Portrait of William Petty as a Doctor of Medicine by the University of Oxford, by Isaac Fuller, c.\ 1649--51, oil on canvas, National Portrait Gallery, London (NPG 2924).  Image courtesy of the National Portrait Gallery.\la{fig:Petty}}
\end{figure*}

Upon his election as a scholar in April 1664, Newton ceased to be a subsizar and his academic career was effectively launched.  According to Westfall, ``the realities of Cambridge in 1664 suggest [...] that Newton had a powerful advocate within the college''.\cite{NaR-patron}  The matter is made all the more puzzling by the lack of a record of active ``networking'' on the part of the young Newton, who was not very socially active and who tended to a morbid aversion of publicity.\cite{NaR-publicity}  An investigation of the identity of the sitter in the Tayler portrait might shed new light on the problem of Newton's early patronage at Cambridge.

Modern historians and biographers have generally downplayed the extent to which Newton may be regarded as Barrow's disciple.\cite{Barrow-Newton}  It is clear, however, that Barrow had a key role in Newton's advancement.  Only five and a half years after Newton's election as a scholar of Trinity, Barrow resigned the Lucasian chair in Newton's favor.  On 26 December, 1671, John Collins reported in a letter that ``Dr Barrow reckons [Newton's unpublished {\it Optical Lectures}] one of the greatest performances of ingenuity this age has afforded''.\cite{Collins1671}  It was Barrow who presented Newton's reflecting telescope to the Royal Society in December 1672, leading to Newton's election as a fellow of that learned body, and later Barrow was one of the first to embrace Newton's theory of colour.\cite{Shapiro}  In 1675, Barrow successfully interceded with King Charles II to allow Newton to remain at Cambridge despite his refusal to take holy orders.\cite{exemption}

%%%%%%%%%%
%%% WILLUGHBY
%%%%%%%%%%
\section{Willughby}
\la{sec:Willughby}

Barrow's Latin edition of Euclid's {\it Elements} is dedicated to three undergraduates: Edward Cecil, John Knatchbull, and Francis Willughby.  These were the wealthiest and most socially eminent students of Trinity College at the time.  Cecil was the sixth son of the Earl of Salisbury, while Knatchbull would later succeed his father as a baronet.  Francis Willughby (1635--1672), the only son of Sir Francis Willughby, was heir to a great estate and a grandson of the Earl of Londonderry.  Unlike most other undergraduates of similar social condition, Willughby was an earnest scholar who completed the full course of university studies (BA, 1656; MA, 1660).\cite{Willughby-education}  Though never an original contributor to mathematics or an active mathematical practitioner, Willughby did show marked interest and aptitude for mathematical studies at Cambridge and Barrow instructed Willughby on the subject.\cite{Willughby-math}

Willughby went on to become one of the original Fellows of the Royal Society in 1663.  In collaboration with John Ray (1627--1705), he made seminal contributions to the systematic study of birds and fish.  An oft-repeated episode connecting him to Newton is that the Royal Society was unable to pay for the printing of Newton's {\it Principia} in 1687 because its budget had been depleted by the publication of the lavishly illustrated {\it De historia piscium} (``On the history of fish'') by Willughby (already deceased) and John Ray, which sold poorly.  Edmond Halley paid for the publication of the {\it Principia} out of his own pocket and was later rewarded by the Royal Society with unsold copies of {\it Historia piscium}.\cite{fishes}

\begin{figure*} [t]
\begin{center}
	\includegraphics[height=0.7 \textwidth]{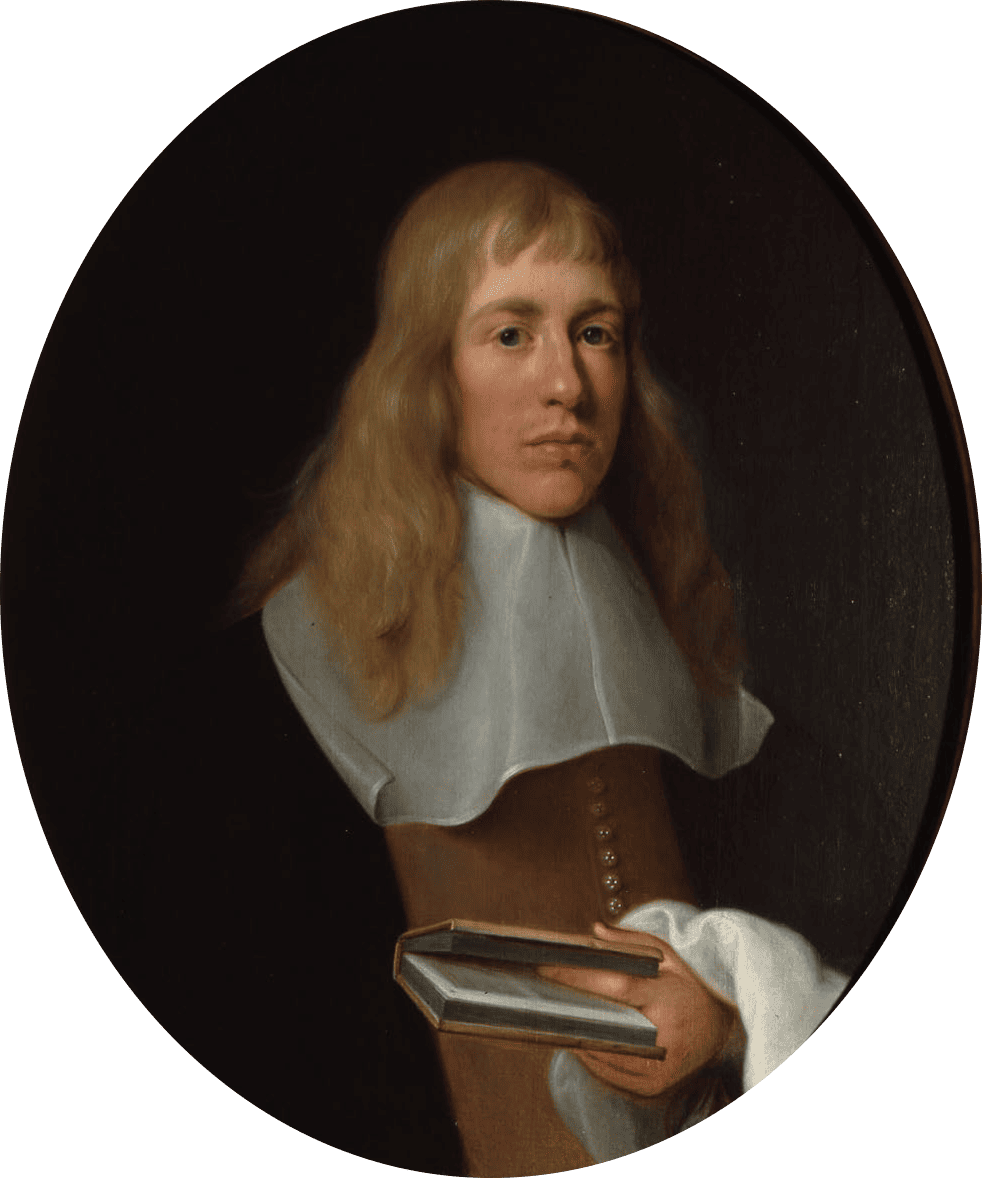}
\end{center}
\caption{Oil portrait of Francis Willughby, c.\ 1652--1660, attributed to Gerard Soest, currently in the private collection of Lord Middleton at Birdsall House, Birdsall, North Yorkshire.  Image courtesy of the Hon.\ James Willoughby and Lady Cara Willoughby.\la{fig:Willughby}}
\end{figure*}

The portrait of Willughby shown in \Fig{fig:Willughby} bears a resemblance to the man in the Tayler picture.  Unlike the Barrow of \Fig{fig:Barrow-Beale}, the Willughby of \Fig{fig:Willughby} shares the blue eyes, high forehead, and straight blond hair of the Tayler's sitter.  It is therefore worth considering that the Tayler could be a portrait of Willughby, shown studying a book dedicated to himself.  I am, however, currently inclined to rank Willughby below Newton as candidate for the sitter.  Their relative social ranks made it appropriate for Barrow to dedicate his first book to Willughby, even though Barrow was an MA and a fellow of Trinity, while Willughby was still an undergraduate.  It therefore seems unlikely that Willughby would have reciprocated the tribute by having his portrait painted with that particular work (in \Fig{fig:Willughby} the book that he is holding is deliberately rendered unidentifiable).  Cecil, Knatchbull, and Willughby, together with Henry Puckering (the son of a baronet and later MP for Warwick) also received in 1660 the dedication of the {\it Homeri Gnomologia} (``Homer's Adages'') by Willughby's tutor at Trinity, the classicist James Duport.\cite{Willughby-education}

Unlike Willughby, the undergraduate Newton would have had an obvious interest in cultivating Barrow's patronage and in publicly presenting himself as Barrow's disciple.  It is also worth noting that Willughby's portrait in \Fig{fig:Willughby}, attributed to the fashionable artist Gerard Soest (c.\ 1600--1681), is obviously of a higher quality and painted in a more modern style than the simpler and more traditional artwork of the Tayler.  This suggests that the painter responsible for the Tayler was a provincial English artist, who would have commanded a significantly smaller fee for his work than a fashionable Dutch-trained artist like Soest.  Given Willughby's wealth and social standing, it seems unlikely to me that he would have also patronized the painter of the Tayler.

The curator of collections at Trinity, Dr.\ Pola Durajska, has suggested to me that Newton might have deliberately used the Tayler picture to set himself up against Willughby as Barrow's true disciple, using the artwork to convey that he was a student capable of deep study and understanding of a book that Barrow had dedicated to Willughby on account of his social position.\cite{Durajska}  This hypothesis seems to offer a plausible interpretation of the origins of the Tayler and its apparent echoes of Soest's portrait of Willughby.

%%%%%%%%%%
%%% BABINGTON
%%%%%%%%%%
\section{Babington}
\la{sec:Babington}

Even though the twice-widowed Hannah had a substantial income that probably exceeded $\pounds 700$ a year, she sent her son Isaac to Trinity in 1661 as a subsizar: one who performed menial tasks for other members of the college, in exchange a partial remission of his fees.  Richard Westfall surmises that Hannah had to be persuaded by the promise of reduced fees to let her eldest son attend university and spare him the obligation of managing the family estate in Lincolnshire.  Two clergymen who had studied at Trinity probably helped convince Hannah to send Isaac to the University.  One was Hannah's brother William Ayscough, the rector of Burton Coggles.  The other was Humphrey Babington, the rector of Boothby Pagnell and a fellow of Trinity, whose sister Mrs.\ Katharine Clark was a close friend of Newton's mother and had boarded the adolescent Newton during his time at the Grantham grammar school.

Babington was personally acquainted with the young Isaac Newton and his mechanical prowess from a very early stage: we know that he supplied the wooden box that the boy used to build a water clock (probably based on instructions from John Bate's {\it Mysteries of Nature and Art}) that his family found useful and which he would leave behind in Woolsthorpe upon his departure for Cambridge.\cite{waterclock}  Newton's friend and biographer, William Stukeley, wrote that Babington ``is said to have had a particular kindness for [Newton] and gave him all the encouragement imaginable, sensible of the lad's great merit.''\cite{kindness}  Westfall speculates that Newton may have gone to Trinity specifically as Babington's servant.\cite{NaR-subsizar}

After he left Grantham for Cambridge, Newton remained friends with Arthur and Katherine Storer, children of Humphrey Babington's sister Katharine by her first husband.  Arthur, who moved to North America at some time before 1673, continued to correspond with Newton on astronomical matters, sometimes with Babington acting as intermediary.  Newton refers in Book III, proposition 41 of the {\it Principia} to Arthur Storer's observations of the Great Comet of 1682, taken near the Patuxent River in the Province of Maryland.\cite{Storer}  When the Great Plague forced the universities to close in the summer of 1665, Newton stayed for a time with Babington at the rectory in Boothby.  Newton later reported that it was there that he had first used the method of power series to compute the area under a hyperbola to 52 decimal places.\cite{Boothby}  There are, therefore, good reasons to suspect that Babington may have taken the fatherless Isaac under his wing during his early years at Cambridge, treating him as a member of his own family.

\begin{figure*} [t]
\begin{center}
	\subfigure[]{\includegraphics[width=0.42 \textwidth]{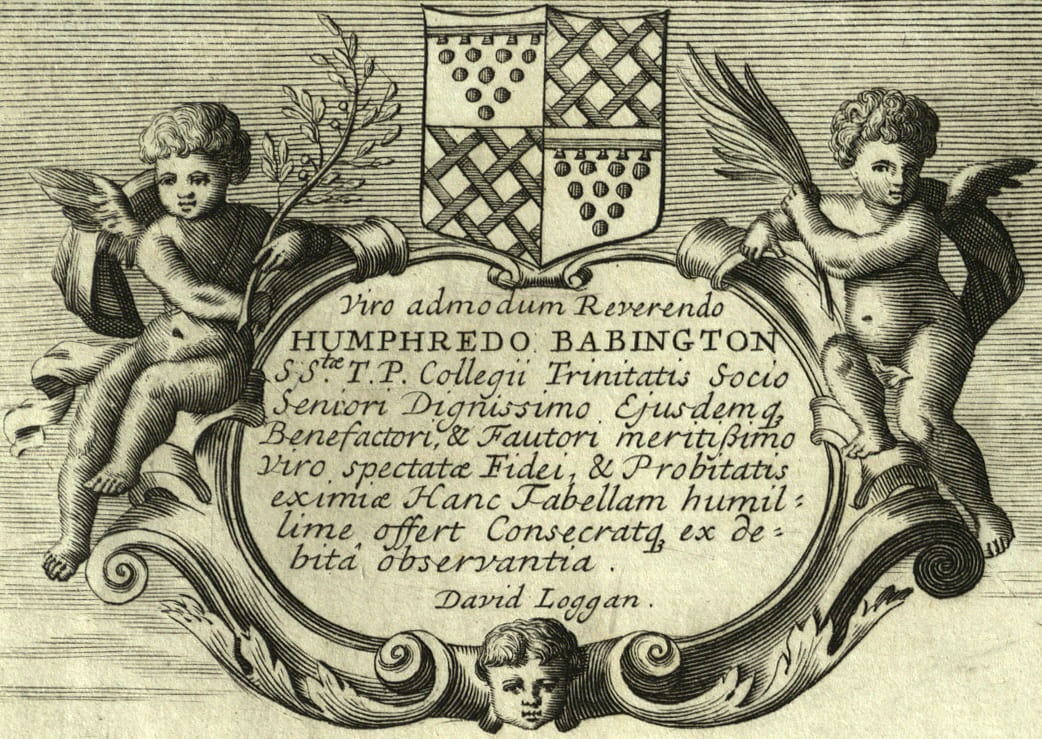}}
	\subfigure[]{\includegraphics[width=0.62 \textwidth]{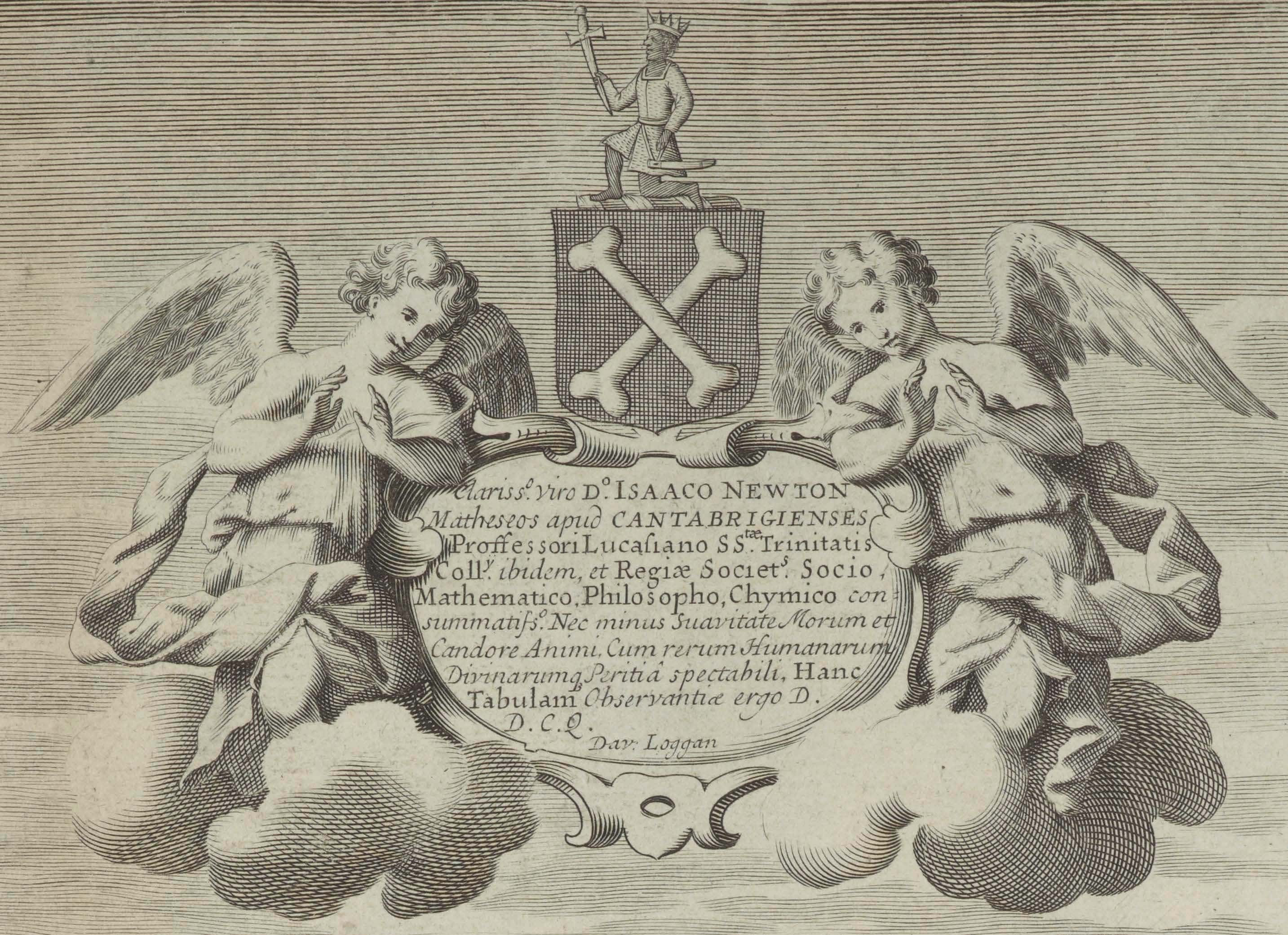}}
\end{center}
\caption{\small (a) Dedication to Humphrey Babington of the first engraved scene (plate V) in David Loggan's {\it Cantabrigia Illustrata}, published c.\ 1690.  (b) Dedication to Isaac Newton of the third scene (plate IX) in the same book.  Images courtesy of the Master and Fellows of Trinity College, Cambridge.\la{fig:Loggan}}
\end{figure*}

Babington was a royalist who had been expelled from Trinity under the Cromwellian Commonwealth, but who returned in triumph after the crown was restored to Charles II.  In 1667 he became a senior fellow of Trinity, one of eight who, together with the Master, governed the college and elected new fellows.  In 1669 he was made a Doctor of Divinity by mandate of the King ({\it per literas regias}).  Later he would serve as Dean, Bursar and, in his final years, Vice-Master of Trinity.  He gave generously to the college and there are some indications that he was a patron of the arts.\cite{Babington-ODNB}

In David Loggan's {\it Cantabrigia Illustrata} (``Cambridge Illustrated''), his remarkable collection of engravings of Cambridge scenes, the very first one (a dual panorama of the town) is dedicated to Babington.  The Latin inscription on this dedication, shown in \Fig{fig:Loggan}(a), can be translated as
\begin{center}
{\it To the most reverend gentleman \\
\textsc{Humphrey Babington}, \\
Doctor of Sacred Theology, most worthy senior Fellow \\
of Trinity College, and its most deserving \\
benefactor and patron; \\
a man of proven faith and outstanding \\
integrity.  This small plate is humbly \\
offered and dedicated to him \\
out of due respect. \\
David Loggan.}
\end{center}
A subsequent scene of Cambridge (a side view of the Church of St Mary the Great) was dedicated to Newton.  The corresponding inscription, shown in \Fig{fig:Loggan}(b), can be translated as
\begin{center}
{\it To the most distinguished man Master \textsc{Isaac Newton}, \\
Lucasian Professor of Mathematics at Cambridge, \\
Fellow of the College of the Most Holy Trinity \\
at the same place, and Fellow of the Royal Society, \\
most accomplished mathematician, philosopher and chymist, \\
no less admirable for his agreeableness of manners \\
and integrity of spirit, together with his skill in things human \\
and divine, David Loggan on account of his veneration \\
gives, dedicates and consecrates this print.}
\end{center}
The fact that this text mentions neither the publication of the {\it Principia} of 1687 nor Newton's election to Parliament in 1689 suggests that the engraving was prepared in the early or mid 1680s, at a time when Newton was virtually unknown beyond a small intellectual circle at Cambridge and in the Royal Society of London.\cite{Illustrata}

The anecdote about Newton's early neglect of Euclid, which we have recounted in \Sec{sec:Euclid}, suggests that Barrow was probably not Newton's {\it first} patron at Trinity and that it might have been necessary for Newton actively to court Barrow's support.  Humphrey Babington did not manifest any personal interest in mathematics or natural philosophy, but he was Newton's friend and protector and understood the usefulness of artistic patronage in advancing a scholar's position in Restoration Cambridge.  Westfall argues that it was probably Babington who secured Newton's election as fellow of Trinity in 1667.\cite{NaR-fellow}  I therefore regard it as plausible that he may have commissioned the Tayler picture ---as well as the dedication of one of the scenes of Loggan's {\it Cantabrigia Illustrata}--- as a favor to his young prot\'eg\'e.

%%%%%%%%%%
%%% CONCLUSIONS
%%%%%%%%%%
\section{Conclusions}
\la{sec:conclusions}

It is my considered judgment that the Tayler picture of \Fig{fig:Tayler}, which unquestionably depicts two pages from Barrow's 1655 Latin edition of Euclid's {\it Elements}, is not of Barrow.  It could be a portrait of Francis Willughby, who was one of three socially preeminent undergraduates at Trinity College to whom Barrow had dedicated that book, and who was a serious student of mathematics.  It could be a portrait of some other young English mathematician who attended university in the 1650s or 1660s, or even somewhat later.  There appear to me several good reasons for taking seriously the possibility that this portrait may be of Isaac Newton, made to mark his election as a scholar of Trinity (on April 1664, aged 21), as fellow (on October 1667, aged 24), or as Barrow's successor to the Lucasian professorship (on October 1669, aged 26).  The last of these options seems disfavored by the Lucasian statutes prescribing scarlet robes, like those of a doctor, for the holder of that chair.\cite{scarlet}  The first option seems to me the most compelling, in light of the sitter's youthful appearance.  Also worth stressing is that Newton's rise from subsizar to scholar of Trinity was the turning point of his meteoric academic career and that we know from Newton's own accounts that his study of Euclid played a key role in it.  

The academic dress depicted in the Tayler picture seems to me consistent with that illustrated by Loggan for Cambridge undergraduates of the later 17th century,\cite{Burgon} but this matter requires further investigation.  Newton became Bachelor of Arts on January 1665 and Master of Arts on July 1668.\cite{Trinity-dates}  A cleaning and re-varnishing of the portrait could clarify whether the sitter's gown was painted as plain black or ``violet'' (as Loggan reports of the gowns of Trinity undergraduates) and whether it is decorated with velvet facings, which could be a sign of a higher status.  Expert examination and restoration of the picture could reveal other clues about its origins and the identities of the artist and sitter.  Whether or not the Tayler turns out to be a portrait of a young Isaac Newton ---surely the most exciting of the possibilities contemplated here--- the matter seems to me worth pursuing.

Tantalizingly little is known about Newton's early life at Cambridge.  Here we have connected the hypothesis that the Tayler picture might depict a young Newton with another hypothesis that was previously advanced by Newton's principal modern biographer, Richard S.\ Westfall: that the fatherless Isaac arrived at the university as a prot\'eg\'e of the Rev.\ Humphrey Babington, a wealthy and well-connected figure in Restoration Cambridge.  Whether these surmises turn out to be correct or not, pursuing them could throw new light on the first stages of one of the most important academic careers in Western history.

%%%%%%%%%%
%%% ACKNOWLEDGEMENTS
%%%%%%%%%%

\begin{acknowledgements}
I thank Dr.\ Nicolas Bell, the librarian of Trinity College, Cambridge, for sharing with me the available information on the provenance of the Tayler portrait (Refs.~\onlinecite{Last-Pye} and \onlinecite{Winstanley}) and for hosting me on a visit to see the original in the Master's Lodge.  I was accompanied on that visit of 1 June 2023 by Dr.\ Vanessa L\'opez Barquero, of the Institute of Astronomy at Cambridge, and Prof.\ Malcolm J.\ Perry, fellow of Trinity.  I thank Dr.\ Pola Durajska, the curator of collections of Trinity, for her encouragement, for the chance to examine the portraits of Figs.\ \ref{fig:Barrow-Beale} and \ref{fig:Seeman-TC} up close, and for her suggestion of how the Tayler picture might have fit into Newton's strategy for gaining patronage (see discussion at the end of \ref{sec:Willughby}).  I also thank Drs.\ Richard Serjeantson of Trinity, and Christopher A.\ Whittaker of Durham University, for their feedback on this manuscript.  The Rev.\ Philip Goff, of the Burgon Society, kindly offered his advice on the academic dress of the relevant period.  Prof.\ Stephen Snobelen, of the University of King's College, Halifax, introduced me to Loggan's {\it Cantabrigia Illustrata} and shared with me the unpublished manuscript in Ref.~\onlinecite{Illustrata}.   

Dr.\ Anne McLaughlin, the Digital Services Manager at Trinity College Library, provided me with the images used in Figs.\ \ref{fig:Barrow-Beale}, \ref{fig:Tayler}, \ref{fig:Seeman-TC}, \ref{fig:Euclid}(a), and \ref{fig:Loggan}.  Dr.\ Colin Higgins, the librarian of St Catharine's College, Cambridge, shared with me the text of Ref.~\onlinecite{Hutton2} and provided the image used in \Fig{fig:Newton-early}(b).  I also thank Ms.\ Gina D'Arcy-Chambers of the Farleigh Wallop Estate, Ms.\ Diana Bularca, digital manager of Philip Mould \& Co., and the Hon.\ James Willoughby for permission to use Figs.\ \ref{fig:Newton-Kneller}, \ref{fig:Newton-early}(b), and \ref{fig:Willughby}, respectively.
\end{acknowledgements}

%%%%%%%%%%
%%% REFERENCES
%%%%%%%%%%

%\bibliographystyle{aipprocl}   	% if natbib is available
%\bibliographystyle{aipprocl} 	% if natbib is missing


\begin{thebibliography}{99}

%%%%%%%%%%
%%% INTRODUCTION
%%%%%%%%%%

\bibitem{Keynes-Garrick}
	M.~Keynes,
	{\it The Iconography of Sir Isaac Newton to 1800}
	(Woodbridge, UK: Boydell P., 2005), pp.\ 3--5
	
\bibitem{Haskell}
	F.~Haskell,
	``The Apotheosis of Newton in Art'', in
	{\it Past and Present in Art and Taste: Selected Essays}
	(New Haven \& London: Yale U.~P., 1987), pp.\ 1--15
	
\bibitem{Manuel-Kneller}
	F.~E.~Manuel,
	{\it A Portrait of Isaac Newton}
	(Cambridge, MA: Harvard U.~P., 1968), p.\ 106
	
\bibitem{Collins1677}
	J.~Collins,
	letter to Isaac Newton dated 5 March 1676/7,
	no.\ 205 in {\it Correspondence of Isaac Newton}, vol.\ III, ed.\ H.~W.~Turnbull
	(Cambridge: Cambridge U.~P., 1961), pp.\ 198--204
	
\bibitem{Houblon}
	A.~Archer Houblon,
	\href{https://archive.org/details/houblonfamilyits02houbuoft/}{{\it The Houblon family, its story and times}, vol.\ II}
	(London: Archibald Constable \& Co., 1907), pp.\ 277--278
	
\bibitem{SirJohn}
The family connection between the Newton baronets and Isaac Newton is detailed in
	Appendix II: Newton's Genealogy,
	in {\it Correspondence of Isaac Newton}, vol.\ VII, eds.\  A.~R.~Hall and L.~Tilling
	(Cambridge: Cambridge U.~P., 1977), pp.\ 485--489.
On the social relations between Sir Isaac and Sir John see, e.g.,
	 A.~R.~Hall,
	{\it Isaac Newton: Adventurer in Thought}
	(Cambridge: Cambridge U.~P., 1992), pp.\ 2, 305, 379.
	
\bibitem{George}
For Col.\ Archer Houblon's genealogy and his connection to the Newton baronets, see Ref.~\onlinecite{Houblon}, pp.\ 315--323.  Some other details on him are given in
	J.~P.,
	``\href{https://doi.org/10.1093/ww/9780199540884.013.U187287}
	{Houblon, Col George Bramston Archer-, (26 June 1843--9 Nov. 1913)}'',
	{\it Who's Who \& Who Was Who}
	(Oxford: Oxford U.~P., 2007).

\bibitem{Keynes-Cook}
Ref.~\onlinecite{Keynes-Garrick}, no.\ XXXVI, pp.\ 52--53

\bibitem{Catz-gentleman}
	H.~Cook,
	``Portrait of a Gentleman (said to be `Sir Isaac Newton, 1642--1727')'',
	oil on canvas (1669), St Catharine's College, Cambridge;
	\url{https://artuk.org/discover/artworks/portrait-of-a-gentleman-194668}
	(accessed 16 Feb.\ 2026)

\bibitem{Hutton1}
	J.~H.~Hutton,
	``Newton and His Portraits'',
	Nature \href{https://doi.org/10.1038/155116c0}{{\bf 155}, 116} (1945)
	
\bibitem{Hutton2}
	J.~H.~Hutton,
	{\it Pictures in the possession of St Catharine's College}
	(Cambridge: Cambridge U.~P., 1950), pp.\ 15--16, plate XXVII
	
\bibitem{Churchill}
	M.~S.~Churchill,
	``{\it The Seven Chapters}, with explanatory notes'',
	Chymia \href{https://doi.org/10.2307/27757273}{{\bf 12}, 29--57} (1967)
	
\bibitem{Reading}
	B.~Reading and P.~Lely,
	``Sir Isaac Newton'' [Image]
	Apollo - University of Cambridge Repository (1799),
	\url{http://www.dspace.cam.ac.uk/handle/1810/218646}
	(accessed 16 Feb.\ 2026)
	
\bibitem{notebook}
	J.~E.~McGuire and M.~Tamny,
	{\it Certain Philosophical Questions: Newton's Trinity Notebook}
	(Cambridge: Cambridge U.~P., 1983)
	
\bibitem{Keynes-Reading}
Ref.~\onlinecite{Keynes-Garrick}, no.\ XXXVIII, pp.\ 53--54

\bibitem{Beale}
	P.~Mould,
	``Portrait of a Mathematician c.\ 1680'',
	in {\it Historical Portraits Picture Archive},
	\url{https://historicalportraits.com/artists/33-mary-beale/works/176-mary-beale-portrait-of-a-mathematician-c.-1680/}
	(accessed 16 Feb.\ 2026)
	
\bibitem{Griffing}
	L.~R.~Griffing,
	``The lost portrait of Robert Hooke?'',
	J.\ Microsc.\ \href{https://doi.org/10.1111/jmi.12828}{{\bf 278}(3), 114--122} (2020)

\bibitem{Whittaker}
	C.~A.~Whittaker,
	``Unconvincing evidence that Beale's {\it Mathematician} is Robert Hooke'',
	J.\ Microsc.\ \href{https://doi.org/10.1111/jmi.12987}{{\bf 282}(2), 189--190} (2021);
	``Has a `lost' portrait of the polymath Robert Hooke (1635-1703) been in London's National Portrait Gallery since 1872?'',
	\url{https://osf.io/preprints/socarxiv/592yf}
	
\bibitem{Hill}
	A.~Hill,
	``Some Account of the Life of Dr.\ Isaac Barrow'',
	in \href{https://archive.org/details/theworksofisaacb01barruoft}{{\it Works of Isaac Barrow}, vol.\ I}
	(New York: John C.\ Riker, 1845 [1683]), pp.\ xi--xxiii.  This includes editorial notes based on Pope's {\it Life of Seth Ward}, Ward's {\it Lives of the Gresham Professors}, and the {\it Biographia Britannica} (see p.\ v).
	
\bibitem{Barrow-Beale}
	M.~Beale,
	``Isaac Barrow (1630--1677), Master (1673--1677), Mathematician and Theologian'',
	oil on canvas, (c.\ 1675), Trinity College, Cambridge (TC Oils P 17);
	\url{https://artuk.org/discover/artworks/isaac-barrow-16301677-master-16731677-mathematician-and-theologian-134861}
	(accessed 16 Feb.\ 2024)

\bibitem{1stDibs}
	M.~Beale,
	``Portrait of Isaac Barrow'',
	oil on canvas (c.\ 1675), private collection; 
	\url{https://www.1stdibs.com/art/paintings/portrait-paintings/mary-beale-portrait-isaac-barrow-17th-century-oil-old-master/id-a_7707102/}		(accessed 16 Feb.\ 2026)

\bibitem{Barrow-Loggan}
	D.~Loggan,
	``Isaac Barrow'',
	plumbago on vellum (1676), National Portrait Gallery, London (NPG 1876);
	\url{https://www.npg.org.uk/collections/search/portrait/mw09078/Isaac-Barrow}
	(accessed 16 Feb.\ 2026)
	
\bibitem{Ward}
	J.~Ward,
	\href{https://archive.org/details/b30450676}{\it Lives of the Gresham Professors}
	(London: J.\ Moore, 1740), p.\ 163
	
\bibitem{Loggan}
Compare the frontispiece of
	I.~Barrow,
	\href{https://archive.org/details/barrowevilspeaking/page/n11/mode/2up}
	{\it Several Sermons Against Evil-Speaking}
	(London: B.\ Aylmer, 1678), in 8vo format and
which is reproduced in \Fig{fig:Barrow-Loggan}(b), and the frontispiece of
	\href{https://archive.org/details/bim_early-english-books-1641-1700_the-works-of-the-learned_barrow-isaac_1683_1}
	{\it The Works of the Learned Isaac Barrow}, vol.\ I
	(London: M.~Flesher, 1683), in 4to format.
The inscription on the latter is mentioned in
	D.~Piper,
	{\it Catalogue of Seventeenth-Century Portraits in the National Portrait Gallery 1625--1714}
	(Cambridge: Cambridge U.~P., 1963), p.\ 20.
	
%%%%%%%%%%
%%% THE TAYLER PICTURE
%%%%%%%%%%
	
\bibitem{Tayler}
	Unknown artist,
	``Isaac Barrow (1630--1677), Master (1673--1677), Mathematician and Theologian'',
	oil on canvas, Trinity College, Cambridge (TC Oils P 16);
	\url{https://artuk.org/discover/artworks/isaac-barrow-16301677-master-16731677-mathematician-and-theologian-134859}
	(accessed 16 Feb.\ 2026)
	
\bibitem{Last}
	L.~Worms,
	``G. H. Last'',
	{\it Antiquarian Booksellers' Association} (2012)
	\url{https://aba.org.uk/page/g-h-last} (accessed 16 Feb.\ 2026);
	``George Herbert Last (1877-1949)'',
	{\it The Bookhunter on Safari} (2012)
	\url{https://ashrarebooks.com/2012/08/20/george-herbert-last-1877-1949/}
	(accessed 16 Feb.\ 2026)
	
\bibitem{CBTayler}
On the older Rev.\ Tayler's career as a popular clergyman and author whose ``numerous books and tracts, which were published on both sides on the Atlantic, were either warnings against the errors of the Catholics or relentlessly moralizing manuals of religious instruction for the young'', see
	G.~C.~Boase and T.~Adams,
	``Tayler, Charles Benjamin (1797--1875)'', in
	{\it Oxford Dictionary of National Biography} (2004)
	\url{https://doi.org/10.1093/ref:odnb/27011}.
	
\bibitem{Last-Pye}
	G.~H.~Last,
	letter to D.~R.~Pye, 26 May 1926.  Trinity College Library, Add.ms.a.232/15
	
\bibitem{Maxwell}
	L.~Campbell and W.~Garnett,
	{\it The Life of James Clerk Maxwell}
	(London: MacMillan, 1882),
	\href{https://archive.org/details/lifejamesclerkm00garngoog/page/169/mode/1up}{pp.\ 169--174, 187--190, 344--345}
	
\bibitem{Keynes-personality}
	M.~Keynes,
	``The personality of Isaac Newton'',
	Notes Rec.\ R.\ Soc.\ Lond.\ \href{https://doi.org/10.1098/rsnr.1995.0001}{{\bf 49}(1), 1--59} (1995)

\bibitem{Winstanley}
	D.~A.~Winstanley,
	``The Portrait of Dr.\ Isaac Barrow'',
	Trinity magazine {\bf 6}(3), 50 (Jun.\ 1925)
	
\bibitem{Seeman-NPG}
	Studio of E.~Seeman,
	``Sir Isaac Newton'',
	oil on canvas (c.\ 1726--1730), National Portrait Gallery, London (NPG 558);
	\url{https://www.npg.org.uk/collections/search/portrait/mw04661/Sir-Isaac-Newton} (accessed 16 Feb.\ 2026)

\bibitem{Keynes-blue}
Ref.~\onlinecite{Keynes-Garrick}, p.\  51

\bibitem{Atterbury}
	F.~Atterbury,
	letter to N.-C.~Thieriot (no.\ LXXVII), in
	\href{https://archive.org/details/epistolarycorre01attegoog}{{\it Epistolary Correspondence}, vol. I}, ed.\ J.~Nichols
	(London: C.~Dilly, 1783), pp.\ 179-182
	
\bibitem{More}
Ref.~\onlinecite{Manuel-Kneller}, p.\ 107
	
\bibitem{Seeman-Chandra}
	S.~Chandrasekhar,
	{\it Newton's} Principia {\it for the Common Reader}
	(Oxford: Clarendon P., 1995), p.\ 302
	
\bibitem{Seeman-Christies}
	E.~Seeman,
	``Portrait of Sir Isaac Newton, Kt. (1642-1727)'',
	oil on canvas (c.\ 1726--1730), private collection
	\url{https://www.christies.com/lot/enoch-seeman-danzig-circa-1694-1744-london-and-5975445} (accessed 16 Feb.\ 2026)

\bibitem{Seeman-Keynes}
Ref.~\onlinecite{Keynes-Garrick}, no.\ XIV, pp.\ 32--35

\bibitem{Seeman-TC}
	E.~Seeman,
	``Isaac Newton (1642--1727), Fellow, Natural Philosopher and Mathematician'',
	oil on canvas (1726), Trinity College, Cambridge (TC Oils P 138);
	\url{https://artuk.org/discover/artworks/isaac-newton-16421727-fellow-natural-philosopher-and-mathematician-134808}
	(accessed 16 Feb.\ 2026)
	
\bibitem{NaR-hair}
	R.~S.~Westfall,
	{\it Never at Rest: A Biography of Isaac Newton}
	(Cambridge: Cambridge U.~P., 1980), p.\ 196

\bibitem{Wickins}
Our source for this is second-hand: it was repeated by John Wickin's son Nicholas in a letter to Prof.\ Robert Smith, dated 16 January 1727/28.  Keynes Ms.\ 137, King's College, Cambridge.  \url{https://www.newtonproject.ox.ac.uk/view/texts/normalized/THEM00035} (accessed 16 Feb.\ 2026)

\bibitem{Fatio}
	K.~Figala and U.~Petzold,
	``Physics and Poetry: Fatio de Duillier's {\it Ecloga} on Newton's {\it Principia}'',
	Arch.\ Int.\ Hist.\ Sci.\ {\bf 37}, 316--349 (1987)
	
\bibitem{alchemy}
	W.~Newman,
	{\it Newton the Alchemist}
	(Princeton and Oxford: Princeton U.~P., 2020), ch.\ 17
	
\bibitem{sophic}
See, e.g.,
	N.~Fatio de Duillier,
	letter to Newton dated 4 May 1693,
	no.\ 414 in {\it Correspondence of Isaac Newton}, vol.\ III, ed.\ H.~W.~Turnbull
	(Cambridge: Cambridge U.~P., 1961), pp.\ 265--267.
The original is item (c) in Keynes Ms.\ 96, King's College, Cambridge.

%%%%%%%%%%
%%% EUCLID, BARROW & NEWTON
%%%%%%%%%%
	
\bibitem{Feingold-meane}
	M.~Feingold,
	``Isaac Barrow: divine, scholar, mathematician'', in
	{\it Before Newton: The Life and Times of Isaac Barrow}, ed.\ M.~Feingold
	(Cambridge: Cambridge U.~P., 1990), p.\ 44
	
\bibitem{Feingold-dates}
On the chronology of publication and travel, see Ref.~\onlinecite{Feingold-meane}, pp.\ 42--43, 47, 52.

\bibitem{NaR-Euclid}
John Conduitt and Antonio Conti independently recorded having heard this story from Newton.  See Ref.~\onlinecite{NaR-hair}, pp.\ 102--103.

\bibitem{H581}
	J.~Harrison,
	{\it The Library of Isaac Newton},
	(Cambridge: Cambridge U.\ P., 1978), item no.\ 581, pp.\ 19, 44, 51, 139, 271.  Harrison describes this item as bearing ``copious annotations by Newton, esp.\ in Books II, V, VII, X; a few signs of dog-earing'' (p.\ 139).  It is currently in the Trinity College library, shelfmark NQ.16.201[1].  High-quality images of all pages are available at \url{https://mss-cat.trin.cam.ac.uk/manuscripts/uv/view.php?n=NQ.16.201}
	(accessed 16 Feb.\ 2026)
	
\bibitem{Whiteside}
The nature of these annotations is discussed briefly in
	D.~T.~Whiteside,
	{\it The Mathematical Papers of Isaac Newton}, vol.\ I (1664--1666)
	(Cambridge: Cambridge U.~P., 1967), p.\ 12.
See also
	S.~Mandelbrote
	{\it Footprints of the lion: Isaac Newton at work}
	(Cambridge: Cambridge U.~P., 2001), item 4, pp.\ 20--22.
	
\bibitem{Wardhaugh}
	B.~Wardhaugh,
	{\it Encounters with Euclid}
	(Princeton and Oxford: Princeton U.~P., 2020), pp.\ 230--236
	
\bibitem{dog-ear}
For a general discussion of Newton's practice of dog-earing his books, see
	J.~Harrison,
	{\it The Library of Isaac Newton},
	(Cambridge: Cambridge U.\ P., 2008 [1978]), pp.\ 25--27.
	
\bibitem{Deanalysi}
On Newton's early work on power series in {\it De analysi per \ae quationes numero terminorum infinitas} (``On the Analysis by means of equations of an infinite number of terms'') and his use of Euclid's method of exhaustion ``to develop an elementary, and too restrictive, convergence test'', see
	N.~Guicciardini,
	{\it Isaac Newton on Mathematical Certainty and Method}
	(Cambridge, MA: MIT Press, 2009), ch.\ 7.

\bibitem{calculus}
On Barrow's proof of the inverse relation between the tangent and quadrature problems, and its influence upon Newton's later work on the calculus, see Ref.~\onlinecite{Deanalysi}, sec.\ 8.1.  At least one important modern textbook refers to the fundamental theorem of calculus as ``Barrow's theorem'':
	V.~I.~Arnol'd,
	{\it Ordinary Differential Equations}, 3rd ed., trans.\ R.~Cooke,
	(Berlin: Springer-Verlag, 1992 [1984]), pp.\ 17--21.
Arnol'd writes (in a modern mixture of Newton's and Leibniz's notation) that the solution to the simplest differential equation, $\dot x = v(t)$, is given by ``Barrow's formula'' as $ x(t) = x_0 + \int_{t_0}^t v(\tau) \, d \tau$.  See also
	M.~Nauenberg,
	``Barrow, Leibniz and the Geometrical Proof of the Fundamental Theorem of the Calculus'',
	Ann.\ Sci.\ \href{https://doi.org/10.1080/00033790.2013.836850}{{\bf 71}(3), 335--354} (2014) [arXiv:1111.6145 [math.HO]].
	
\bibitem{Bentley}
	R.~Bentley,
	``Paper of directions given by Newton to Bentley respecting the books to be read before endeavouring to read and understand the {\it Principia}'',
	no.\ 367 in {\it Correspondence of Isaac Newton}, vol.\ III, ed.\  H.~W.~Turnbull
	(Cambridge: Cambridge U.~P., 1977), pp.\ 155-156

\bibitem{Principia-LG}
	C.~R.~Leedham-Green,
	``Translator's Preface'', in
	I.~Newton,
	{\it The Mathematical Principles of Natural Philosophy}
	(Cambridge: Cambridge U.\ P., 2022), p.\ xii
	
\bibitem{mixed}
	D.~Levitin,
	{\it The Kingdom of Darkness: Bayle, Newton, and the Emancipation of the European Mind from Philosophy}
	(Cambridge: Cambridge U.~P., 2022), pp.\ 509--525

\bibitem{Petty-portrait}
	I.~Fuller,
	``Sir William Petty'',
	oil on canvas (c.\ 1649--1650), National Portrait Gallery, London (NPG 2924);
	\url{https://artuk.org/discover/artworks/sir-william-petty-155656}
	(accessed 16 Feb.\ 2026)
	
\bibitem{Piper-Petty}
	D.~Piper,
	{\it Catalogue of Seventeenth-Century Portraits in the National Portrait Gallery 1625--1714}
	(Cambridge: Cambridge U.~P., 1963), pp.\ 275--276
	
\bibitem{Petty}
	T.~Barnard,
	``Petty, Sir William (1623--1687)'', in
	{\it Oxford Dictionary of National Biography} (2004),
	\url{https://doi.org/10.1093/ref:odnb/22069}

\bibitem{NaR-patron}
Ref.~\onlinecite{NaR-hair}, p.\ 102

\bibitem{NaR-publicity}
See, e.g., Ref.~\onlinecite{NaR-hair}, pp.\ 224--226.

\bibitem{Barrow-Newton}
See
	M.~Feingold,
	``Newton, Leibniz, and Barrow Too: An Attempt at a Reinterpretation'',
	Isis \href{https://doi.org/10.1086/356464}{{\bf 84}(2), 310--38} (1993),
and references therein.

\bibitem{Collins1671}
	J.~Collins,
	letter of 26 Dec.\ 1671 to F.~Vernon and R.~Towneley, in
	\href{https://archive.org/details/MathematicsIsaacNewtonVol3_1670-73Whiteside1969/MathematicsIsaacNewtonVol3_1670-73Whiteside1969_144x75/}
	{{\it The Mathematical Papers of Isaac Newton}, vol. 3}, ed.\ D.~T.~Whiteside
	(Cambridge: Cambridge U.~P., 1969), p.\ 23
	
\bibitem{Shapiro}
	A.~E.~Shapiro,
	``The Gradual Acceptance of Newton's Theory of Light and Color'',
	Perspect.\ Sci.\ \href{https://doi.org/10.1162/posc_a_00498}{{\bf 4}(1), 59--140} (1996)

\bibitem{exemption}
According to Ref.~\onlinecite{NaR-hair}, p.\ 333, ``we know nothing of the factors behind the recorded events'' of this royal dispensation, but ``it appears more likely that on this occasion it was Isaac Barrow who rescued Newton from threatened oblivion''.  More recently, Levitin declares flatly that ``via Barrow, Newton secured letters patent (issued under the broad seal on 27 April 1675) to exempt the Lucasian Professor from this requirement'' of ordination; Ref.~\onlinecite{mixed}, p.\ 544.  See also
	S.~Mandelbrote
	{\it Footprints of the lion: Isaac Newton at work}
	(Cambridge: Cambridge U.~P., 2001), item 5, pp.\ 22--23.
	
%%%%%%%%%%
%%% WILLUGHBY
%%%%%%%%%%
	
\bibitem{Willughby-education}
	R.~Serjeantson,
	``The Education of Francis Willughby'', in
	{\it Virtuoso by Nature: The Scientific Worlds of Francis Willughby FRS (1635--1672)},
	ed.\ T.~Birkhead
	(Leiden: Brill, 2016), pp.\ 44--98
	
\bibitem{Willughby-math}
	B.~Wardhaugh,
	``Willughby's Mathematics'', in
	{\it Virtuoso by Nature: The Scientific Worlds of Francis Willughby FRS (1635--1672)},
	ed.\ T.~Birkhead
	(Leiden: Brill, 2016), pp.\ 122--141

\bibitem{fishes}
On the anecdote about publication of the {\it Principia}, see Ref.~\onlinecite{NaR-hair}, p.\ 453.  On the work of Willughby and Ray, see 
	S.~Kusukawa,
	``The {\it Historia Piscium} (1686)'',
	Notes Rec.\ R.\ Soc.\ Lond.\
	\href{https://doi.org/10.1098/rsnr.2000.0106}{{\bf 54}(2), 179--197} (2000);
	``{\it Historia Piscium} (1686) and Its Sources'', in
	{\it Virtuoso by Nature: The Scientific Worlds of Francis Willughby FRS (1635--1672)},
	ed.\ T.~Birkhead
	(Leiden: Brill, 2016), pp.\ 305--334.
	
\bibitem{Durajska}
	P.~Durajska,
	private communication with the author (2025).
According to Serjeantson, it was ``more in his capacity as an armigerous fellow-commoner than as a budding mathematician [\ldots] that Willughby received in 1655 the dedication of Isaac Barrow's concise Latin version of Euclid's {\it Elements} (Ref.~\onlinecite{Willughby-education}, p.\ 48).

%%%%%%%%%%
%%% BABINGTON
%%%%%%%%%%
	
\bibitem{waterclock}
	W.~Stukeley,
	``Memoirs of Sir Isaac Newton's life'',
	in {\it Early Biographies of Isaac Newton 1660--1885}, vol.\ 1, ed.\ R.~Iliffe
	(London \& New York: Routledge, 2016 [1752]), pp.\ 269--270
	
\bibitem{kindness}
Ref.~\onlinecite{waterclock}, p.\ 279

\bibitem{NaR-subsizar}
Ref.~\onlinecite{NaR-hair}, pp.\ 72--74

\bibitem{Storer}
	P.~Broughton,
	``Arthur Storer of Maryland: His Astronomical Work and his Family Ties with Newton'',
	J.\ Hist.\ Astron.\ \href{https://doi.org/10.1177/002182868801900201}{{\bf 19}(2), 77--96} (1988)
	
\bibitem{Boothby}
	 A.~R.~Hall,
	{\it Isaac Newton: Adventurer in Thought}
	(Cambridge: Cambridge U.~P., 1992), pp.\ 2, 19, 31, 33, 54
	
\bibitem{Babington-ODNB}
	J.~Parkin,
	``Babington, Humphrey (1615--1692)'', in
	{\it Oxford Dictionary of National Biography} (2004),
	\url{https://doi.org//10.1093/ref:odnb/974}

\bibitem{Illustrata}
	D.~Loggan,
	{\it Cantabrigia Illustrata}
	(Cambridge: printed by the author c.\ 1690), plates IV, IX.
High resolution scans of this book are available at \url{https://luna.folger.edu/luna/servlet} (accessed 16 Feb.\ 2026).  For a modern edition with English translation, introduction, and notes see
	J.~W.~Clark (ed.),
	Cantabrigia Illustrata {\it by David Loggan}
	(Cambridge: Macmillan \& Bowes, 1905).
I have also consulted
	S.~D.~Snobelen,
	``David Loggan's dedication to Isaac Newton on his engraving of Great St. Mary's Church, Cambridge (1690)'', unpublished manuscript.
	
\bibitem{NaR-fellow}
Ref.~\onlinecite{NaR-hair}, pp.\ 177--178

%%%%%%%%%%
%%% CONCLUSIONS
%%%%%%%%%%
	
\bibitem{scarlet}
Ref.~\onlinecite{NaR-hair}, p.\ 208.  Although he does not say so explicitly in Refs.\ \onlinecite{Hutton1} or \onlinecite{Hutton2}, Hutton's conclusion that Fig.~\ref{fig:Newton-early}(b) was a portrait of Newton as the new Lucasian Professor might have been influenced by the red robe shown there.

\bibitem{Burgon}
	B.~M.~Newman,
	``The Evolution of Undergraduate Academic Dress at the University of Cambridge and its Constituent Colleges'',
	Trans.\ Burgon Soc.\ \href{https://doi.org/10.4148/2475-7799.1189}{{\bf 20}(1) 67--93} (2021)

\bibitem{Trinity-dates}
These dates are given in Ref.~\onlinecite{NaR-hair}, chs.\ 3, 6.
	
\end{thebibliography}
\end{document}